\documentclass[preprint]{aastex}
\usepackage{mkfig}
\begin{document}

%\title{Image-Based Reconstruction of the CMEs that Hit Parker Solar Probe
%  on 2021 April 24-25}
%\title{Parker Solar Probe Imaging and In~situ Observations of a Series of
%  Small Coronal Mass Ejections}
\title{Resolving Red Giant Winds with the
  {\em Hubble Space Telescope}\altaffilmark{1}}

\author{Brian E. Wood\altaffilmark{2}, Graham M. Harper\altaffilmark{3},
  Hans-Reinhard M\"{u}ller\altaffilmark{4}}
\altaffiltext{1}{Based on observations made with the NASA/ESA Hubble
  Space Telescope, obtained at the Space Telescope Science Institute,
  which is operated by the Association of Universities for Research
  in Astronomy, Inc., under NASA contract NAS 5-26555.  These
  observations are associated with programs GO-15903 and `GO-15904.}
%\altaffiltext{1}{DISTRIBUTION A.  Approved for public release:
%  distribution unlimited}
\altaffiltext{2}{Naval Research Laboratory, Space Science Division,
  Washington, DC 20375, USA; brian.e.wood26.civ@us.navy.mil}
\altaffiltext{3}{CASA, University of Colorado, Boulder, CO 80309-0389, USA}
\altaffiltext{4}{Department of Physics and Astronomy, Dartmouth College,
  Hanover, NH 03755, USA}
%\altaffiltext{2}{George Mason University, 4400 University Drive, Fairfax, VA
%  22030, USA}
%\altaffiltext{3}{The Johns Hopkins University Applied Physics Laboratory,
%  Laurel, MD 20723, USA}

%\linenumbers

\begin{abstract}

     We describe recent spectroscopic observations of red giant stars
made by the Space Telescope Imaging Spectrograph (STIS) instrument on
board the {\em Hubble Space Telescope}, which have provided
spatially resolved observations of the warm chromospheric winds
that predominate for early K to mid-M giants.  The
H~I Lyman-$\alpha$ lines of a set of 11 red giants observed with
the STIS/E140M echelle grating are first analyzed to
ascertain wind H~I column densities and total wind mass-loss rates.
The M giants have estimated mass-loss rates of
$\dot{M}=(14-86)\times 10^{-11}$ M$_{\odot}$~yr$^{-1}$, while the K giants
with detected wind absorption have weaker winds with
$\dot{M}=(1.5-2.8)\times 10^{-11}$ M$_{\odot}$~yr$^{-1}$.  We
use long-slit spectra of H~I Lyman-$\alpha$ for two particular
red giants, $\alpha$~Tau (K5~III) and $\gamma$~Cru (M3.5~III), to study
the spatial extent of the Lyman-$\alpha$ emission.  From these data
we estimate limits for the extent of detectable
emission, which are $r=193$~R$_*$ for $\gamma$~Cru and $r=44$~R$_*$
for $\alpha$~Tau.  Cross-dispersion emission profiles
in the STIS echelle spectra of the larger sample of
red giants also show evidence for spatial resolution, not only
for H~I Lyman-$\alpha$ but for other lines with visible wind
absorption, such as Fe~II, Mg~II, Mg~I, O~I, and C~II.
We characterize the nature of these spatial signatures.  The spatial
extent is far more apparent for the M giants than for the K giants,
consistent with the stronger winds found for the M giants from the
Lyman-$\alpha$ analysis.

\end{abstract}


\section{Introduction}

     The winds of red giant stars with spectral types between K2~III and
M5~III represent an interesting transition between the hot, fast,
relatively weak winds of coronal stars like the Sun
($\dot{M}_{\odot}=2\times 10^{-14}$ M$_{\odot}$ yr$^{-1}$), and the cool,
very slow, very massive winds of M supergiants and pulsating M giants
(e.g., Mira variables), with $\dot{M}\sim 10^{-7}$ M$_{\odot}$ yr$^{-1}$.
The winds of K and early M giants are typically found to have velocities
of $20-40$ km~s$^{-1}$, with mass loss rates typically estimated at
$\dot{M}=10^{-11}$ to $10^{-10}$ M$_{\odot}$~yr$^{-1}$, with warm
temperatures of $T\sim 10^{4}$~K \citep[e.g.,][]{amb79,eog13,gr18,gmh22}.
The origin of the winds is unclear, but acceleration
by Alfv\'{e}n waves is one possibility, which would require the presence
of magnetic fields \citep{lh80,tks07,va10}.
Although signatures of the winds are often detectable in the
optical Ca~II H \& K lines \citep{dr77}, the best diagnostics of the
winds are in the UV, where the winds produce broad absorption
troughs in strong chromospheric resonance lines like H~I Lyman-$\alpha$
$\lambda$1216, O~I $\lambda 1300$, C~II $\lambda 1335$,
Mg~II h \& k $\lambda\lambda2796, 2803$, and numerous near-UV (NUV)
Fe~II lines \citep{gmh95,rdr98}.

     The UV spectroscopic capabilities of the {\em Hubble Space Telescope}
(HST) have made it much easier to study red giant winds, particularly the
higher resolution capabilities of the current Space Telescope Imaging
Spectrograph (STIS) instrument \citep{rak98,bew98},
and the former Goddard High Resolution Spectrograph (GHRS)
instrument.  Prior to HST, the state-of-the-art were spectra from
the {\em International Ultraviolet Explorer} (IUE).  Not only is
signal-to-noise (S/N) much improved with HST, but the higher spectral
resolution makes it much easier to separate stellar wind
absorption from ISM absorption in strong lines like Mg~II h \& k.
The point spread function (PSF) of HST is also much better
characterized.

     As described by \citet[][hereafter Paper 1]{bew16}, we have
conducted an HST spectroscopic survey of nine carefully selected red giant
stars in order to provide a consistent database of spectra for studying
the chromospheric emissions from these stars, absorption from their
stellar winds, and absorption from the wind-ISM interaction regions
(i.e., their ``astrospheres''). The HST/STIS observations of these
stars, which were made between 2013 October and 2015 January,
consisted of both high-resolution
($R\equiv \lambda/\Delta\lambda = 110,000$) spectra of the NUV
2574--2851~\AA\ wavelength with the E230H grating, and a longer
exposure of the far-UV (FUV) 1150--1700~\AA\ wavelength range taken
with the moderate resolution ($R=46,000$) E140M grating.  In Paper~1,
the observations of the nine survey targets were supplemented by
archival HST observations of four other red giants, including two
particularly bright and nearby stars, $\alpha$~Tau (K5~III) and
$\gamma$~Cru (M3.5~III), which are of particular interest here.

     Paper~1 focused almost entirely on the Mg~II h \& k lines,
demonstrating clear correlations between spectral type and Mg~II flux, 
Mg~II k/h flux ratio, and inferred wind speed.  This is unlike
main sequence stars, for which chromospheric emission is far more
correlated with stellar rotation rate.  However, red giants are
almost uniformly very slow rotators, and the strong spectral type
correlations noted above are evidence that the chromospheres of
all red giants are emitting at the same minimum ``basal'' flux level.
The chromospheric wind speeds are likewise highly correlated with
spectral type.

     Astrospheric absorption was also a focus of Paper~1.  Supersonic
stellar winds expand radially from a star until the wind pressure
decreases to a value equivalent to the surrounding ISM pressure, at
which point the stellar wind experiences a termination shock (TS)
where the wind is heated, compressed, and decelerated to a subsonic
speed.  For a red giant wind, this is hundreds of au from the star
\citep{bew07}.  For the Paper~1 sample of stars, absorption
from this post-TS wind material was detected in the Mg~II lines of
three stars:  $\alpha$~Tau (K5~III), $\sigma$~Pup (K5~III), and
$\gamma$~Eri (M1~III).  Such absorption represents an independent
diagnostic of stellar wind strength, as well as a unique remote
probe of ISM properties and termination shock physics.  We note here
in passing that for $\gamma$~Eri a follow-up HST observation was
made to try to detect the star's astrosphere in Mg~II h \& k emission,
as the detected astrospheric Mg~II absorption feature should be
indicative of photon scattering and not pure absorption.
On 2020 February 23, the star was imaged with the NUV F275W filter of
the Wide Field Camera 3 (WFC3) instrument, but unfortunately the
astrospheric scattered-light Mg~II emission was not detected.

     No attempt was made to estimate mass loss rates from Mg~II in
Paper 1, partly because such estimates ideally involve radiative
transfer modeling, which was outside the scope of the paper, but
also because mass loss rate estimates should ideally consider
observations of other spectral lines besides Mg~II.  Of particular
importance is the H~I Lyman-$\alpha$ line, which samples the dominant
mass constituent of the stellar wind, hydrogen.  One goal of
this paper will be to measure wind H~I column densities from the
STIS/E140M spectra of Lyman-$\alpha$, which provide direct estimates
of mass loss rates.  In the course of this Lyman-$\alpha$ analysis,
we serendipitously discovered that the Lyman-$\alpha$ emission seen
in the STIS/E140M echelle spectra is actually spatially resolved,
especially for the M giants in our sample.

     The STIS/E140M observations were obtained using the
narrow $0.2^{\prime\prime}\times 0.2^{\prime\prime}$ aperture.
Such data are not intended to provide spatial information, but
cross-dispersion emission profiles nevertheless can be used to
distinguish between a point source and extended emission that is
filling the aperture more uniformly.  Inspection of the
cross-dispersion profiles reveal clear spatial resolution of the
emission over a broad wavelength region surrounding the wind and
ISM absorption, particularly for the M giants.  We also looked at
the cross-dispersion profiles of other lines in the STIS E230H and E140M
data, and we once again found evidence for spatial resolution in
other chromospheric lines with wind absorption features (e.g., Mg~II,
Mg~I, Fe~II, O~I, C~II), albeit over a narrower wavelength range.
A second goal of this paper will be to provide a thorough
empirical study of the spatial extent of the emission based on
the cross-dispersion E230H and E140M emission profiles for our
broad sample of red giant stars.

     Finally, for two red giants, $\alpha$~Tau (K5~III) and
$\gamma$~Cru (M3.5~III), we obtained STIS long-slit FUV spectra of
the H~I Lyman-$\alpha$ spectral region using the G140M grating,
in order to provide observations that can provide more detailed
information about the full spatial extent of the Lyman-$\alpha$
emission, at least along one direction.  The two chosen stars are
not technically among our nine survey stars, but they were chosen for
their brightness and close proximity.  A third goal of the paper will
be to present and discuss these more recent long-slit spectral
observations.

     The STIS E140M/E230H cross-dispersion data for our survey targets
and the STIS/G140M long-slit observations for $\alpha$~Tau and
$\gamma$~Cru represent the first spatially resolved observations of
red giant winds in the UV, and therefore represent an important step
forward for our understanding of these winds.
%A fourth and final goal of this
%paper will be to provide an initial attempt to model the UV observations
%using radiative transfer codes, focusing on $\gamma$~Cru.  This
%modeling will also consider constraints provided by radio data.

\begin{deluxetable}{lccccccccccc}
\tabletypesize{\scriptsize}
\tablecaption{HST Red Giant Targets}
\tablecolumns{12}
\tablewidth{0pt}
\tablehead{
  \colhead{Star} & \colhead{Alternate} & \colhead{Spect.} &
    \colhead{R$_*$\tablenotemark{a}} & \colhead{d\tablenotemark{b}} &
    \colhead{$\phi$\tablenotemark{c}} & \colhead{$V_w$\tablenotemark{d}} &
    \colhead{$\log N_H$\tablenotemark{e}} &
    \colhead{$\log N_H$\tablenotemark{f}} &
    \colhead{$\log F_H$\tablenotemark{g}} &
    \colhead{$\log F_H$\tablenotemark{h}} &
    \colhead{$\dot{M}$\tablenotemark{i}} \\
  \colhead{} & \colhead{Name} & \colhead{Type} &
    \colhead{(R$_{\odot}$)} & \colhead{(pc)} &
    \colhead{(mas)} & \colhead{(km/s)} &
    \colhead{wind} &
    \colhead{ISM} &
    \colhead{observed} &
    \colhead{corrected} &
    \colhead{($10^{-11}$)} }
\startdata
\multicolumn{12}{l}{\underline{Original Survey Targets}} \\
HD 66141  & HR 3145     & K2 III  & 23.9 & 77.9 & 2.85 &(42)&$<18.4$& 18.56 &
  4.52 & 4.86 & $<1.8$ \\
HD 211416 & $\alpha$~Tuc& K3 III  & 37.3 & 61.2 & 5.66 &(43)&$<18.6$& 18.79 &
  4.36 & 4.82 & $<4.7$ \\
HD 87837  & HR 3980     & K4 III  & 33.6 & 90.6 & 3.45 &(36)&$<19.3$& 19.53 &
  3.48 & 4.34 & $<18$ \\
HD 50778  & $\theta$~CMa& K4 III  & 35.4 & 79.9 & 4.12 & 30 & 18.34 & 17.97 &
  4.35 & 4.64 & 1.7 \\
HD 59717  & $\sigma$~Pup& K5 III  & 43.7 & 59.4 & 6.83 & 43 & 18.05 & 18.87 &
  4.32 & 4.72 & 1.5 \\
HD 25025  & $\gamma$~Eri& M1 III  & 58.9 & 62.3 & 8.78 & 24 & 19.12 & ...   &
  4.15 & 4.59 & 14 \\
HD 44478  & $\mu$~Gem   & M3 III  &107.7 & 71.0 & 14.1 & 19 & 19.76 & ...   &
  3.27 & 4.01 & 86 \\
HD 20720  & $\tau^4$~Eri& M3.5 III&102.9 & 93.4 & 10.2 & 23 & 19.65 & ...   &
  3.41 & 4.13 & 78 \\
HD 120323 & HR 5192     & M4.5 III& 82.4 & 56.1 & 13.6 & 19 & 19.79 & ...   &
  3.08 & 4.04 & 70 \\
\multicolumn{12}{l}{\underline{Other Stars}} \\
HD 29139  & $\alpha$~Tau& K5 III  & 51.0 & 20.4 & 23.2 & 35 & 18.33 & 18.29 &
  4.17 & 4.49 & 2.8 \\
HD 108903 & $\gamma$~Cru& M3.5 III& 73.9 & 27.2 & 25.2 & 28 & 19.52 & ...   &
  3.52 & 4.17 & 50 \\
\enddata
\tablenotetext{a}{Stellar radius.}
\tablenotetext{b}{Stellar distance.}
\tablenotetext{c}{Angular diameter.}
\tablenotetext{d}{Wind terminal speeds from Mg~II spectra, with values
  in parentheses assumed rather than measured (see text).}
\tablenotetext{e}{Wind H~I column density (in cm$^{-2}$).}
\tablenotetext{f}{ISM H~I column density (in cm$^{-2}$).}
\tablenotetext{g}{Directly observed logarithmic H~I Lyman-$\alpha$
  surface flux (in ergs cm$^{-2}$ s$^{-1}$).}
\tablenotetext{h}{Absorption corrected logarithmic H~I Lyman-$\alpha$
  surface flux (in ergs cm$^{-2}$ s$^{-1}$).}
\tablenotetext{i}{Mass loss rate inferred from wind N$_H$ (in 10$^{-11}$
  M$_{\odot}$~yr$^{-1}$).}
\end{deluxetable}

\section{Stellar Wind H~I Column Densities and Inferred Mass-Loss Rates}

     Table~1 provides a list of the red giant target stars considered
here, very similar to the sample from Paper~1, which provides more
information about where some of the basic stellar information was
obtained.  The table includes stellar wind terminal velocities
($V_w$) measured from the Mg~II lines in Paper~1.  For the first three
stars in the table no such measurements are available so we list values
computed from the empirical $V_w$--$T_{eff}$ relation from Paper~1.

\begin{figure}[t]
\plotfiddle{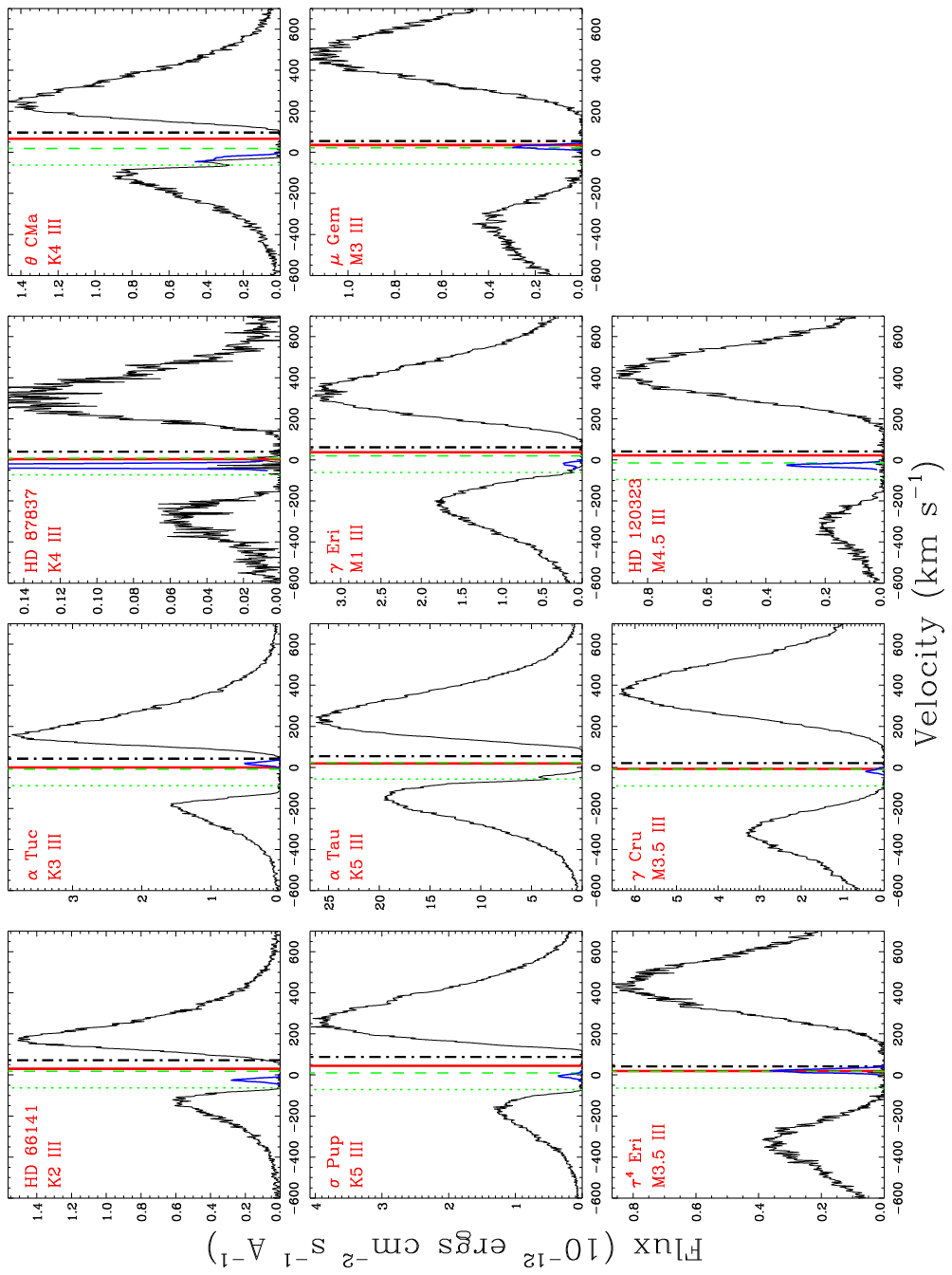}{3.9in}{-90}{70}{70}{-285}{360}
\caption{STIS/E140M spectra of the H~I Lyman-$\alpha$ line, plotted
  on a heliocentric velocity scale.  The red giant spectra are shown in
  order of spectral type.  The narrow blue emission is geocoronal
  emission (too weak to be seen for $\alpha$~Tau).  The vertical
  dot-dash line is the stellar rest frame.  The vertical red line is
  the terminal wind velocity, measured from Mg~II data in Paper~1.
  The vertical green dashed and dotted lines are the expected locations
  of ISM H~I and D~I absorption.}
\end{figure}
     STIS/E140M spectra are available for all the stars, providing
high quality spectra of the H~I Lyman-$\alpha$ line at 1215.67~\AA.
The Lyman-$\alpha$ spectra are shown in Figure~1, which
displays the data in order of spectral type, from K2~III to
M4.5~III.  Broad, saturated absorption is seen at the center of the
chromospheric emission, which is nominally a blend
of absorption from the ISM and absorption from the stellar wind.
The narrow, blue emission peaks are geocoronal emission, which in
all but one case is fully contained within the saturated core
of the absorption, and is therefore trivial to remove.  The exception
is $\theta$~CMa, where the geocoronal emission is blended with
the blue side of the absorption profile.  Even for this case it is
possible to remove the emission, guided by the width of the emission
seen for the other stars, and by knowledge of the emission's expected
central velocity (e.g., the component of the Earth's velocity towards
the observed line of sight).

     Vertical dot-dashed lines in Figure~1 indicate the stellar rest
frame, based on known radial velocities of the stars (see Paper~1).
Vertical red lines indicate the wind velocities ($V_w$) from Table~1.
Green vertical lines indicate the expected location of the ISM H~I and
D~I (deuterium) absorption, based on the Local Interstellar Cloud (LIC)
vector of \citet{sr08}.  Paper~1 demonstrated that the
Mg~II spectra of our target stars have absorption at the expected
LIC velocity.  Additional ISM velocity components are also apparent
in some of the Mg~II spectra, which is not uncommon even for very
nearby lines of sight, but these extra components are not generally
widely separated from the LIC velocity.

     The primary goal here is to estimate stellar mass loss rates
($\dot{M}$) from the H~I Lyman-$\alpha$ wind absorption.  There are
several advantages of estimating $\dot{M}$ from Lyman-$\alpha$
instead of from wind absorption in other lines such as Mg~II or
Fe~II, as is more commonly done \citep{gr18,gmh95,gmh22}.
One is that hydrogen is the dominant
mass constituent of the stellar wind, meaning that an $\dot{M}$
measurement from Lyman-$\alpha$ will not be reliant on assumptions
about elemental abundances, though it will still be reliant on
assumptions about hydrogen's uncertain ionization state in the wind.
A second is that the Lyman-$\alpha$ absorption is so broad that
details of the wind velocity structure are unimportant.
For all the stars in Figure~1, the
absorption is much broader than the velocity difference between the
stellar rest frame and the wind terminal velocity.
Finally, H~I column densities are high enough that the absorption is
in the damping regime of the curve of growth, particularly
for the stronger M giant winds in Figure~1.  This means that
the Doppler core of the absorption, defined by wind temperature
and/or turbulent velocity, is relatively unimportant, and the
absorption profile is defined solely by the line-of-sight integrated
H~I column density through the wind to the star, $N_H$ (in units of
cm$^{-2}$), which is the primary quantity of interest.

     The main complication with measuring $N_H$ for our stars
is separating the wind absorption from ISM absorption.  In
making this separation we rely on a long history of analyzing ISM
Lyman-$\alpha$ absorption observed towards nearby stars
\citep{bew05,bew21}.  This experience provides familiarity
with what the ISM absorption should look like towards our red giant
targets.  All of these targets were chosen to lie within the
bounds of the Local Bubble (LB), which is a broad region lying
within roughly 100~pc from the Sun in most directions, where the
ISM has generally low density and is in fact mostly hot and ionized
\citep{dms99,rl03,cz22}.
There are some warm, partially neutral clouds within the LB, and it
so happens that the Sun lives within one of them, namely the
aforementioned LIC.

     An interstellar line's opacity profile, $\tau_{\lambda}$,
is a Voigt profile that depends on three parameters:
column density (N, in cm$^{-2}$), central velocity
($V$), and a Doppler broadening
parameter ($b$).  The Doppler parameter includes
both thermal and nonthermal velocities, added in quadrature.
For Lyman-$\alpha$, both H~I and D~I absorption must be
considered, with the rest wavelength of D~I Lyman-$\alpha$
lying $-0.33$~\AA\ from H~I Lyman-$\alpha$.
We naturally assume that the central velocities of the
absorbing H~I and D~I are the same.  For the ISM within the LB,
the Doppler broadening is dominated by thermal motions, which
means that we can assume $b_D=b_H/\sqrt{2}$.  Finally, it has
been found that within the LB, the D/H abundance ratio is uniform,
${\rm D/H}=1.56\times 10^{-5}$ \citep{bew04}, so we
can simply assume this ratio in our analysis.  In this way, the
D parameters are fully connected to the H parameters, and a
full H+D Lyman-$\alpha$ opacity profile depends on only the
three H parameters.  The transmission profile is then
$\exp{(-\tau_{\lambda})}$, which is multiplied by the background
emission profile to provide an absorption profile for comparision
with observations.  Extensive examples of these kinds of
analyses can be found elsewhere \citep[e.g.,][]{bew05,bew21}.

     A major advantage of our analysis is that we are not just
studying one or two stars, which individually could have anomalous
wind or line-of-sight ISM properties.  Confidence is increased by
seeing similar behavior from similar types of stars.
For example, Paper~1 clearly showed how Mg~II wind absorption
profiles change with spectral type, strengthening as spectral
type changes from the K giants to the M giants.  This pattern
is clear despite occasional confusion with ISM absorption for
certain lines of sight.  The empirical Mg~II spectra by
themselves imply that $\dot{M}$ increases with
spectral type, although no $\dot{M}$ measurements were actually
made in Paper~1.  In Figure~1, we clearly see that the
Lyman-$\alpha$ absorption generally broadens with spectral type,
with the sides of the absorption becoming shallower in slope.
This is also consistent with $\dot{M}$ increasing with spectral
type.  The one obvious anomaly is HD~87837, a K giant showing
very broad, strong absorption.  Is this due to an unexpectedly
high ISM column density, or unexpectedly strong wind absorption?
We favor an ISM origin, as the Mg~II spectra presented in
Paper~1 certainly do not suggest a strong wind for this star,
but more on the subject will be said below.

\begin{figure}[t]
\plotfiddle{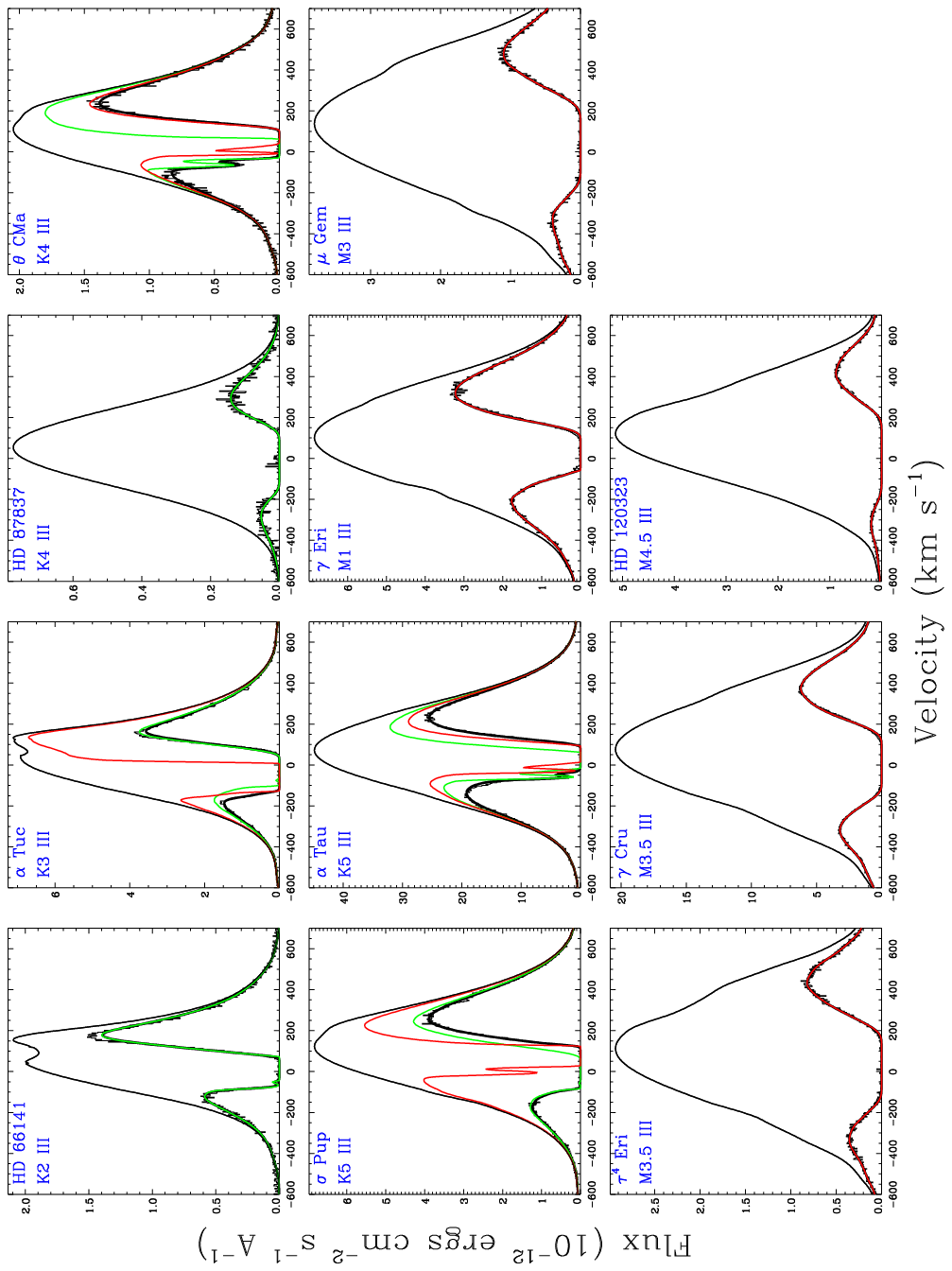}{3.9in}{-90}{70}{70}{-285}{360}
\caption{Fits to the Lyman-$\alpha$ data from Figure~1, where the
  upper solid line is the reconstructed stellar emission line.
  The absorption is modeled as either ISM absorption (green lines)
  or stellar wind absorption (red lines), or a combination of
  the two.}
\end{figure}
     For the M giants in the sample, a preliminary analysis
finds that the absorption profiles all imply $\log N_H>19.0$.
Interstellar column densities this high are very rare within the
LB, with the only example of which we are aware being HD~82558,
with $\log N_H=19.05$ \citep{bew05}.  For this reason, we
are reasonably confident that the absorption seen for the five
M giants in Figure~1 can be assumed to be entirely from the
stellar wind.  For these stars, we fit the data assuming a simple
Gaussian for the stellar line profile and a single wind
absorption profile, treating the wind opacity profile in the same
way that we would an ISM opacity profile, with the same three free
parameters, keeping in mind that the
entire wind velocity structure is buried in the core of the broad
absorption profile and is therefore unimportant.  We assume the
same D/H ratio in the winds as are using for the ISM, though at
the high column densities of the M giant winds, the D absorption
is irrelevant, as it is completely obscured by the
broad, saturated H absorption.  After
an initial fit, the assumed background stellar profile is modified
to improve the quality of fit, and then a second, final fit is
performed.  The final fits and assumed background line profiles
are shown in Figure~2.  The inferred wind column densities are
listed in Table~1, falling in the range $\log N_H=19.12-19.79$
for the M giants.

     We have fitted the unexpectedly strong absorption seen
towards HD~87837 in a manner identical to the M giants, finding
a column density of $\log N_H=19.53$ (see Figure~2).  This is a
much higher column density than we would normally expect to see
for the ISM within the LB, but we nevertheless conclude that
this is indeed ISM absorption.  One clue for what is going on
lies in an alternate name for the star, 31~Leo, indicating
a star in the constellation of Leo.  Perhaps
the most anomalous ISM structure known within the LB is
the Local Leo Cold Cloud (LLCC), which is a remarkably cold
($T\approx 20$~K) and dense ($n_H\approx 3000$~cm$^{-3}$)
ISM structure within the LB, perhaps about 20~pc away,
first detected in H~I 21~cm emission \citep{jegp11,dmm12}.
The observations suggest that the LLCC is
a sheet-like structure that is very thin, perhaps only
$\sim 200$~au thick.  This could represent a collision between
ISM flows, and this collisional interface extends well beyond
the bounds of the cold LLCC, perhaps even including the LIC
itself \citep{cg17,ps22}.  In any case,
our red giant, HD~87837, happens to lie roughly behind the LLCC,
explaining the anomalously high ISM column density.
Our $\log N_H=19.53$ measurement is to our knowledge
the highest column density measured from Lyman-$\alpha$ for a
star within the LB, implying an average density of
$n_H=0.12$~cm$^{-3}$ over the 90.6~pc line of sight to the star.

     For the other three K4-K5 giants in our sample, we believe the
Lyman-$\alpha$ absorption is a blend of absorption from the ISM
and the stellar wind.  For $\theta$~CMa and $\alpha$~Tau, the
Lyman-$\alpha$ analysis is helped greatly by the clear presence
of separated ISM D~I absorption at the expected LIC velocity
(see Figure~1).  This means that the D~I absorption can be used to
constrain the ISM H~I absorption.  The ISM cannot account for the
extent of the absorption on the red side of the H~I absorption profile,
which requires wind absorption to explain.
We fit the data in a similar manner as the M giants, but with an
ISM absorption component in addition to the wind component.
The resulting two-component fits are shown in Figure~2.  For
$\sigma$~Pup, there is no clearly distinct ISM D~I absorption feature,
but the blueward extent of the observed absorption matches well with
the expected location of the LIC D~I absorption, and like
$\theta$~CMa and $\alpha$~Tau, it is not possible for the ISM
to account for the absorption on the red side of the line,
suggesting that wind absorption is dominating that side of the
absorption.  The $\sigma$~Pup data are therefore also modeled with
a two-component fit (see Figure~2).  The $\log N_H$ wind and ISM
measurements resulting from these K giant
analyses are listed in Table~1.  Uncertainties for the K giant
measurements will naturally be higher than for the M giants
given the difficult separation between wind and ISM absorption.

     There are two remaining Lyman-$\alpha$ spectra left to be
described.  For HD~66141, we find that the observed absorption can
be acceptably fit with only ISM absorption, with no evidence for
any stellar wind contribution at all (see Figure~2).  The
$\alpha$~Tuc Lyman-$\alpha$ data are much harder to explain.
The difficulty is apparent in Figure~1, with both the ISM
and wind H~I absorption expected to be centered well to
the red of the actual observed center of the H~I absorption.
There does not seem to be an easy way for any model to explain
the blueward extent of the observed absorption profile.

     We hypothesize that this is evidence for a high-velocity
wind component for $\alpha$~Tuc, analogous to the high-velocity
($V_w=74$ km~s$^{-1}$) component observed for the hybrid
chromosphere star $\gamma$~Dra in Mg~II spectra (see Paper~1).
Hybrid chromosphere stars are red giants with both strong
chromospheric winds and also high temperature emission from lines
such as C~IV $\lambda$1548.  It seems likely that the high speed
wind component seen for $\gamma$~Dra is associated with the
atmospheric regions producing the star's high temperature emission.
There is no detected high velocity wind component for $\alpha$~Tuc
in Mg~II, but perhaps such a wind does exist, but with insufficient
Mg~II column density for detection in the h \& k lines.
Thus, the two-component fit shown for the $\alpha$~Tuc data in
Figure~2 includes not only an ISM component but also a high-velocity
wind component forced to be centered at a wavelength corresponding
to $V_w=100$ km~s$^{-1}$.  This high velocity wind absorption
allows us to account for the blue side of the absorption
profile.  However, because of the tentative nature of this wind
detection, we do not report a wind column density for this
component in Table~1.  For this star, and for the two other red
giants without wind measurements (HD~66141, HD~87837), we instead
provide upper limits for the chromospheric wind column density
in Table~1.

\begin{figure}[t]
%\plotfiddle{fig3.ps}{3.5in}{-90}{63}{63}{-250}{325}
\plotfiddle{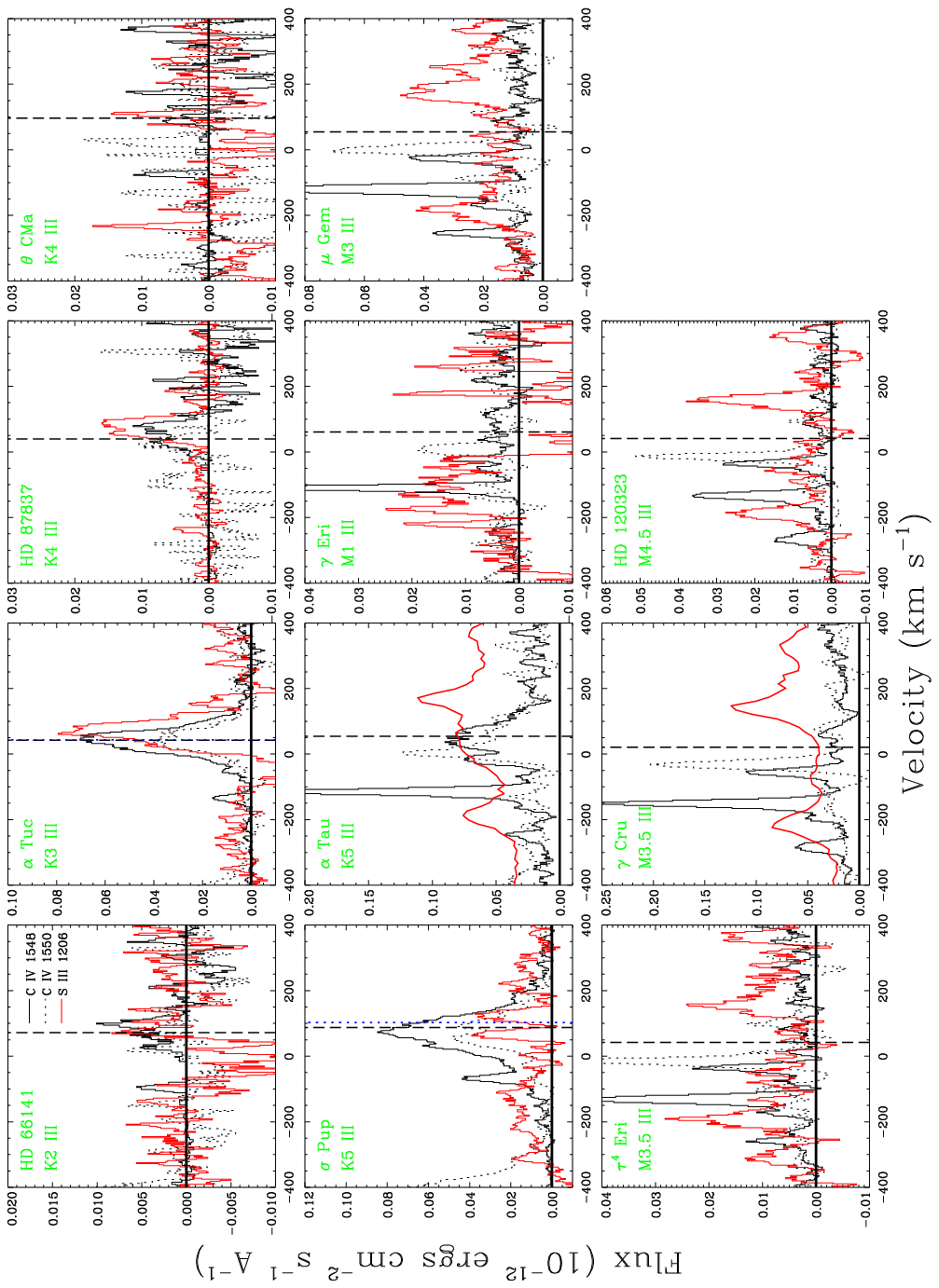}{3.2in}{-90}{60}{60}{-250}{310}
\caption{STIS/E140M spectral regions containing the
  C~IV $\lambda$1548 (solid black), C~IV $\lambda$1550 (dotted black),
  and Si~III $\lambda$1206 (solid red) lines, plotted with significant
  smoothing on a heliocentric velocity scale.  Most spectra show
  no evidence of the high temperature emission centered at the
  systemic radial velocity (vertical black dot-dashed lines).
  For $\alpha$~Tuc and $\sigma$~Pup, which are spectroscopic binaries,
  vertical blue dotted lines indicate the actual rest frame of the red
  giant primary based on H$_2$ lines.  Only $\alpha$~Tau, $\alpha$~Tuc,
  and $\sigma$~Pup show clear detections of C~IV and/or Si~III, but for
  $\alpha$~Tuc and $\sigma$~Pup the emission may be from the companion
  stars and not the red giants.  Narrow chromospheric lines are seen
  in many of the spectra, particularly for the M giants.  For the
  Si~III spectra of $\alpha$~Tau and $\gamma$~Cru, the E140M data have
  been replaced with the higher S/N G140M spectra described in Section~4.}
\end{figure}
     Our case for $\alpha$~Tuc being a star analogous to
$\gamma$~Dra would be strengthened by evidence for high temperature
line emission.  The STIS/E140M observations that provide our H~I
Lyman-$\alpha$ spectra also cover the wavelengths of many high
temperature lines, the strongest of which are the
C~IV $\lambda\lambda$1548, 1550 doublet and the Si~III line at
1206.5~\AA.  In Figure~3, we show the C~IV and Si~III spectral
regions for all the stars in our sample, not just $\alpha$~Tuc.
For purposes of the figure, the noisy spectra have been smoothed
using a 5-pixel running boxcar to try to reveal faint C~IV and
Si~III emission.  For $\alpha$~Tau and $\gamma$~Cru, the G140M
spectra described below in Section~4 are shown for Si~III instead
of the E140M spectra, due to much higher S/N.

     The clearest C~IV and Si~III emission is seem for
$\alpha$~Tuc and $\sigma$~Pup.  Unfortunately, both of these stars are
spectroscopic binaries, which complicates the situation considerably.
For $\sigma$~Pup at least, it seems likely that the G5~V companion
is responsible for the emission.  As noted in Paper~1, on the
basis of H$_2$ lines the rest frame of the K5~III star at the time
of observation is actually $+103$ km~s$^{-1}$, different from the
systemic $V_{rad}=87.3$ km~s$^{-1}$ value listed in Table~1.
The C~IV emission in Figure~3 is clearly blueshifted relative to
this, and is more centered on the rest frame of the G5~V companion,
suggesting the companion as the more likely source of the emission.

     In contrast, the H$_2$ lines of $\alpha$~Tuc suggest a rest
frame for the K3~III giant of $+43$ km~s$^{-1}$, nearly identical to
the systemic $V_{rad}=42.2$ km~s$^{-1}$ value in Table~1, and
in good agreement with the observed location of the C~IV and Si~III
emission.  The combined C~IV flux observed for $\alpha$~Tuc,
$f=4.7\times 10^{-14}$ ergs~cm$^{-2}$~s$^{-1}$, represents a surface
flux that is about 2.7 times higher than observed for $\gamma$~Dra
\citep{tra06}.  Although higher, it is similar enough for it
to be plausible that this emission is from
$\alpha$~Tuc, providing some support for the hypothesis that
$\alpha$~Tuc is a hybrid chromosphere star.  However, there is still
significant ambiguity about the emission's origins, particularly
because very little is known about $\alpha$~Tuc's companion, making
it difficult to rule it out as the source.

     The Si~III line from $\alpha$~Tuc has a profile that looks
absorbed on the blue side, exactly like what might be expected
from the kind of hot, fast wind we are proposing to explain the
the Lyman-$\alpha$ absorption in Figure~2.  This would be an
exciting finding, but confidence in the reality of this absorption
is low, as the absorption seems to actually dip below zero flux.
Data reduction can be problematic for this spectral region in
STIS echelle data, due to close proximity with the much brighter
H~I Lyman-$\alpha$ line.  Scattered light from
Lyman-$\alpha$ leads to a strong, uneven background for the Si~III
line, which is difficult to properly remove \citep{gmh22}.

     There is clearly C~IV emission apparent from $\alpha$~Tau
in Figure~3, which has been noted and measured previously
\citep{tra03}.  We can now report a detection of Si~III as
well, thanks to the G140M spectrum presented here for the first time
(see Section~4).  The Si~III line is a broad emission feature bracketed
by a narrow emission line on each side.  Not even a hint of this
line is apparent in the E140M spectrum, possibly due in part to the
data reduction issues mentioned above.  Although peripheral
to our analysis here, we can a report a flux for the $\alpha$~Tau
Si~III line of $f=2.8\times 10^{-14}$ ergs~cm$^{-2}$~s$^{-1}$ based
on a multi-Gaussian fit to the G140M data.

     Besides $\alpha$~Tuc, $\sigma$~Pup, and $\alpha$~Tau, the only
other star with some evidence for C~IV or Si~III emission in Figure~3
is HD~87837.  However, the weak emission is not well centered on the
stellar rest frame, making this a dubious identification.
For the M giants, no C~IV or Si~III lines are apparent, but there
are many narrow, low temperature lines observed instead,
presumably Fe~II and S~I \citep{tra03,kgc18}.

\begin{figure}[t]
\plotfiddle{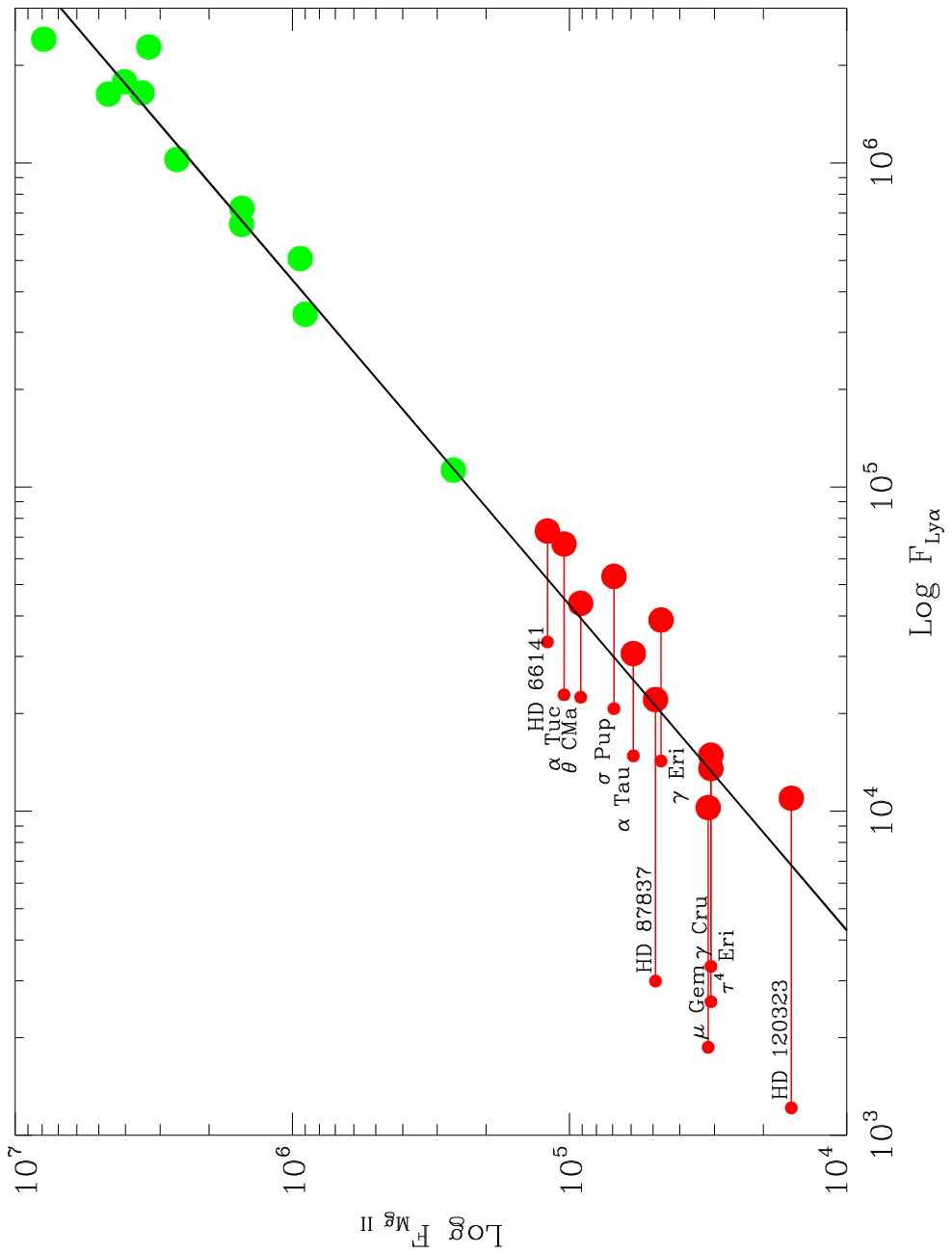}{3.6in}{-90}{63}{63}{-255}{335}
\caption{Mg~II surface fluxes plotted versus H~I Lyman-$\alpha$ surface
  fluxes for giant stars.  The green data points are for coronal
  stars analyzed by Wood et al.\ (2005), and the black line shows a
  linear fit to these data points alone.  The red data points are
  for our sample of later type giants with chromospheric winds.  The
  small red circles indicate the observed Lyman-$\alpha$ flux, while
  the larger red circles indicate the flux from the reconstructed
  stellar Lyman-$\alpha$ line profile from Figure~2.}
\end{figure}
     Returning to the Lyman-$\alpha$ analysis, the Lyman-$\alpha$
lines are important chromospheric diagnostics, so we measure line
fluxes for our target stars.  There are two Lyman-$\alpha$ line fluxes
that can be measured from Figure~2, one simply measured from the
observed line profile, and one for the absorption-corrected stellar
line profiles shown in the figure.  The second flux measurement is
naturally larger than the first.  These fluxes are converted to
surface fluxes and both quantities are reported in Table~1.

     It is instructive to compare the Lyman-$\alpha$ surface fluxes
with the Mg~II h \& k surface fluxes measured in Paper~1.  This
comparison is made in Figure~4.  Both Lyman-$\alpha$ flux values
are shown in the figure, with horizontal lines connecting the
two values for each star.  The Mg~II--H~I relation for our sample
of wind-dominated late-type giants can be compared with that of
coronal giants with earlier spectral types (F8-K1~III).  The
coronal giant measurements shown in Figure~4, and the linear fit
to them, are from Figure~15(b) of \citet{bew05}.

     The later-type ``windy'' giants are reasonably consistent with
the coronal Mg~II--H~I flux-flux relation only if the higher,
absorption-corrected line fluxes are used.  This is unsurprising
for the K giants, with Lyman-$\alpha$ absorption that is
ISM-dominated, and which therefore requires a correction for
the Lyman-$\alpha$ photons absorbed by the ISM.  However, Figure~4
suggests that this is also true for the M giants.  This is unexpected,
as these Lyman-$\alpha$ lines are clearly dominated by stellar wind
rather than ISM absorption.

     Although our analysis has treated the Lyman-$\alpha$ absorption
like true absorption features regardless
of whether the absorption is ISM-dominated or wind-dominated,
the stellar winds are in truth generally assumed to be scattering
photons, not destroying them.  It is generally assumed
that photons are being scattered from line center into the line wings,
where opacities are low enough for the photons to finally escape.
Thus, the true H~I Lyman-$\alpha$ flux for a wind scattering dominated
line profile should simply be the observed integrated line flux, and
not that inferred from the kind of reconstructed line profiles shown
in Figure~2, which are implicitly assuming that photons are being
destroyed rather than just scattered.  Figure~4 seems to instead
suggest that the higher absorption-corrected line fluxes are the
correct fluxes.  This would imply that Lyman-$\alpha$ photons
are actually being destroyed somehow, and not simply scattered.
Either that is true or the M giant chromospheres are intrinsically
deficient in Lyman-$\alpha$ emisison.  With hotter and cooler
chromospheric plasma being mixed for these red giants
\citep[e.g.,][]{tra03}, one possible mechanism for
photon destruction is via the bound-free continuum of C~I.

     Assuming a $1/r^2$ falloff in density, mass loss rates can
be computed using the following equation:
\begin{equation}
\dot{M}=\frac{4\pi m_p r_{\circ} V_w N_H}{\chi_1 \chi_2},
\end{equation}
where $m_p$ is the proton mass, $V_w$ is the wind velocity, $N_H$ is
the wind H~I column density, $\chi_1$ is the fraction of the
wind mass that is hydrogen, $\chi_2$ is the fraction of hydrogen
in the wind that is neutral, and $r_{\circ}$ is a reference radius
for the base of the absorption region.  This equation assumes a
wind that accelerates quickly, with the bulk of the absorption
therefore at the wind terminal velocity.  The $\chi_1$ factor
accounts for the mass contribution of elements other than H to the
wind, primarily He.  Based on solar abundances, we estimate
$\chi_1=0.75$ \citep{ma09}.  The $\chi_2$ factor
corrects for H ionization.  There is evidence for significant
ionization of H in red giant winds \citep[e.g.,][]{gmh22},
so we estimate $\chi_2=0.5$.  The $V_w$ and $N_H$ values for our stars
are listed in Table~1, the former deriving from the Mg~II study in
Paper~1 and the latter from the Lyman-$\alpha$ analysis performed here.
We make the usual assumption that the chromospheric absorption is
coming from close to the star, and simply assume $r_{\circ}=1.2~R_*$,
where the stellar radii, $R_*$, are provided in Table~1.
Equation~(1) implicitly assumes a
spherically symmetric wind.  The assumption could lead to either
underestimates or overestimates of $\dot{M}$, depending on whether the line
of sight to the star is sampling parts of the wind with densities lower
or higher than average.

     The resulting $\dot{M}$ measurements for our red giant stars
are listed in the last column of Table~1.  The
$\dot{M}=2.8\times 10^{-11}$ M$_{\odot}$~yr$^{-1}$ value for $\alpha$~Tau
is in reasonable agreement with past measurements based on Mg~II and
Fe~II lines \citep{rdr98,bew07,gr18}.  The $\gamma$~Cru measurement is in
good agreement with that reported by \citet{gr18}, but our
$\mu$~Gem measurement is higher than the \citet{gr18} value by an order
of magnitude.  It should be noted that intrinsic variability
could also contribute to these discrepancies, as red giant winds are
known to exhibit some degree of variability.  Both IUE monitoring of
Mg~II and ground-based monitoring of Ca~II have shown variations in wind
absorption for $\alpha$~Tau and $\gamma$~Cru \citep{wlk78,djm98}.
Furthermore, $\gamma$~Cru was observed by both HST/GHRS and HST/STIS,
showing modest variability.  More specifically, based on Fe~II lines
\citet{ken23} report $V_w=18.6$ km~s$^{-1}$ and
$\dot{M}=4.8\times 10^{-10}$ M$_{\odot}$~yr$^{-1}$ from the GHRS data,
and $V_w=16.2$ km~s$^{-1}$ and
$\dot{M}=3.8\times 10^{-10}$ M$_{\odot}$~yr$^{-1}$ from the STIS data.

\begin{figure}[t]
\plotfiddle{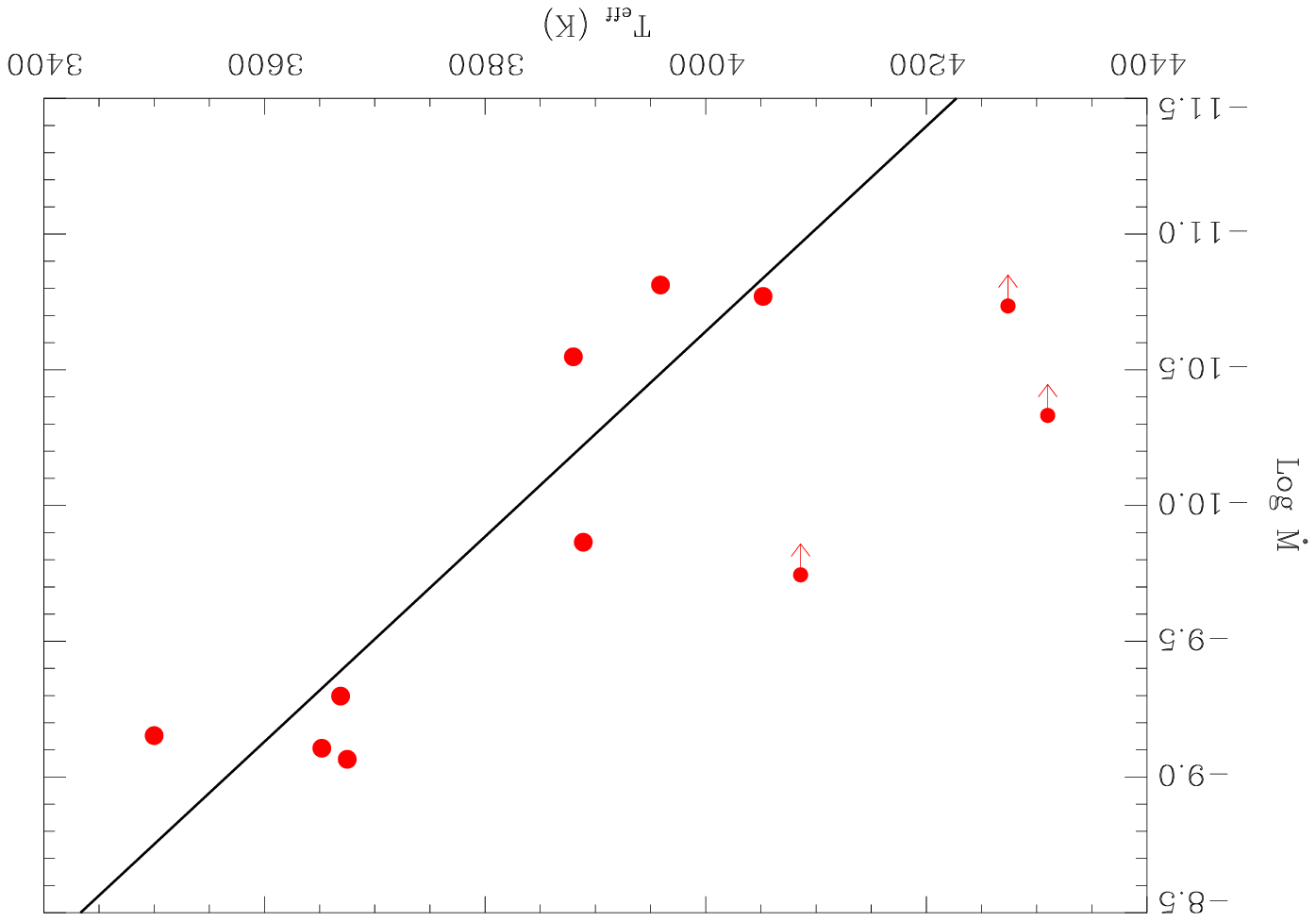}{3.6in}{180}{63}{63}{240}{335}
\caption{Mass loss rates (in M$_{\odot}$~yr$^{-1}$) plotted versus
  stellar photospheric temperature, with a linear fit to the
  data.}
\end{figure}
     The Lyman-$\alpha$ derived $\dot{M}$ values are clearly
correlated with spectral type, with the M giants having significantly
stronger winds.  In Figure~5, $\dot{M}$ is plotted versus the
photospheric effective temperature, following a pattern established
in Paper~1, where the assumed $T_{eff}$ values are listed in that
paper's Table~1.  A linear fit is performed to the relation in
Figure~5.
%, which is
%\begin{equation}
%\log \dot{M}=-32.7058\log T_{eff} + 106.753.
%\end{equation}
It is worth noting that the well-studied K1.5~III giant
$\alpha$~Boo would not be very consistent with this relation, with
the $\dot{M}=(2.5-4.0)\times 10^{-11}$ M$_{\odot}$~yr$^{-1}$
measurement from \citet{gmh22} being about a factor
of 10 higher than Figure~5 would predict.  The $\alpha$~Boo
Mg~II spectrum was discussed in Paper~1, and it does appear to have
significantly stronger wind absorption than other stars of similar
spectral type, at least in our sample of stars.  An even larger
sample than ours would be required to assess the degree to which
$\alpha$~Boo is unusual in terms of its chromosphere.  It does,
however, have unusual kinematics, being the brightest member of
the Arcturus stream, a thick disc population of stars with
uncertain origin \citep[e.g.,][]{ik19}.

\section{Spatial Resolution Inferences from Cross-Dispersion Emission Profiles}

\subsection{The Mg~II h \& k Lines}

     As noted in Section~1, during the Lyman-$\alpha$ analysis described
in Section~2 we noticed that the cross-dispersion profiles of the
STIS/E140M Lyman-$\alpha$ emission are not point-source-like.  Similar
evidence for spatial resolution was found for other
lines.  We first present the evidence for this in the Mg~II h \& k lines
that were the focus of Paper~1.

\begin{figure}[t]
\plotfiddle{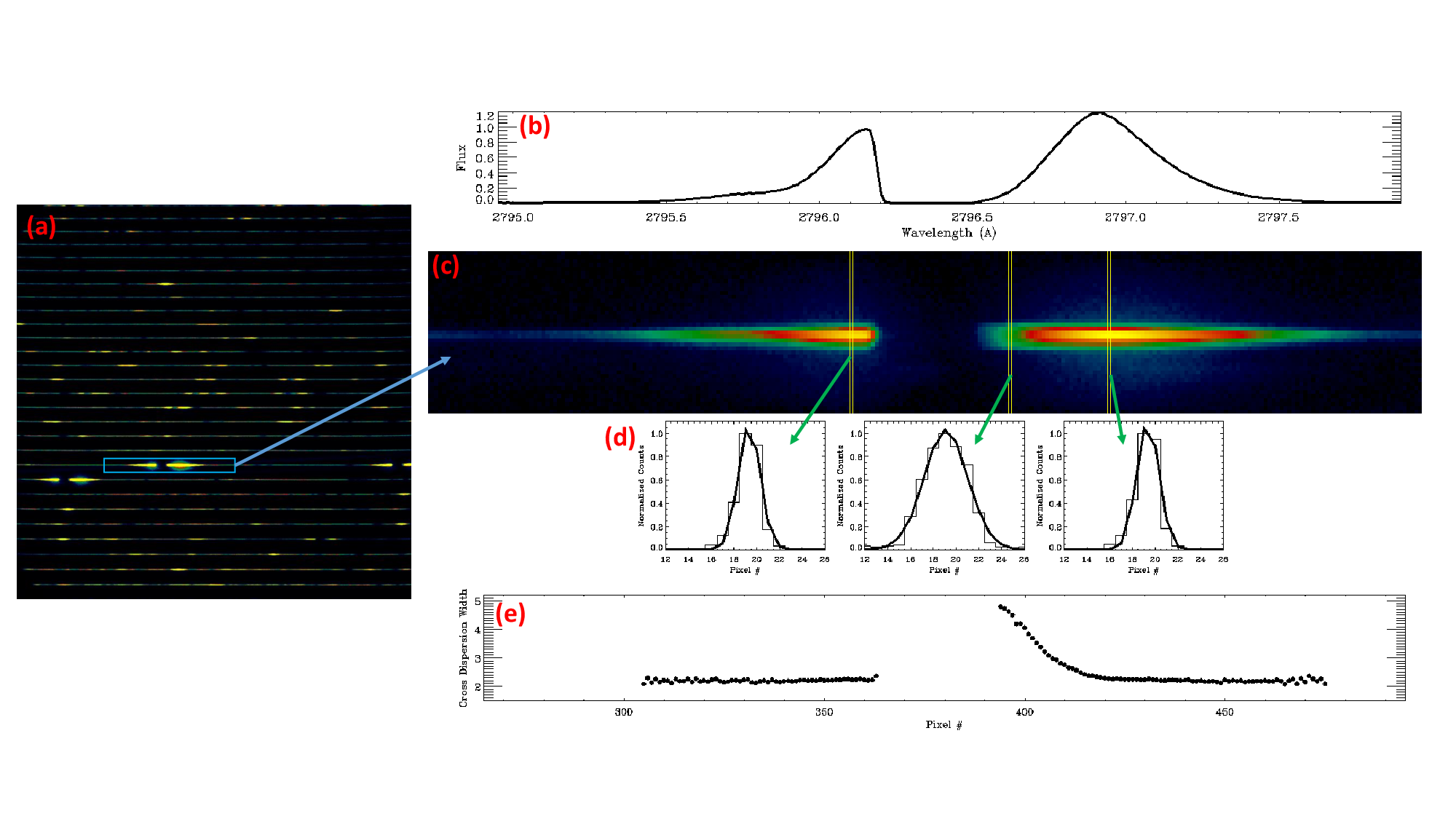}{3.0in}{0}{53}{53}{-245}{-40}
\caption{(a) STIS/E230H observation of $\gamma$~Cru, showing 27 spectral
  orders stacked vertically, covering $2624-2895$~\AA.  The Mg~II k
  line is outlined.  (b) The extracted 1-D Mg~II k line spectrum, with
  flux in units of $10^{-12}$~ergs~cm$^{-2}$~s$^{-1}$~\AA$^{-1}$,
  with the center of the line affected by wind absorption.
  (c) Close-up of the E230H echelle image of the Mg~II k line from
  (a).  (d) Three cross-dispersion profiles for three locations in the
  k line, with the first and third showing no spatial resolution, and
  the second showing significant resolution.  Gaussians are fitted to
  the cross-dispersion profiles.  (e) Cross-dispersion widths (FWHM,
  in pixels) plotted versus pixel \#, with clear spatial resolution
  signatures only apparent along the red side of the wind absorption.}
\end{figure}
     Figure~6(a) shows a raw STIS/E230H observation of $\gamma$~Cru
(Obs.\ ID \#OBKK82020), taken through the
$0.2^{\prime\prime}\times 0.09^{\prime\prime}$ aperture.  In this
image the 27 usable spectral orders are stacked on top of each other,
with wavelength increasing from top to bottom.  The brightest line in
the $2624-2895$~\AA\ spectral region is the Mg~II k line, which is
outlined in the figure.  The Mg~II h line is visible to its right in
the same spectral order.  The wavelength ranges of the orders can
overlap, so the h line is also visible in the next order, below and to
the left of the k line.

     A blow-up of the k line from panel (a) is shown in Figure~6(c), and
above this in Figure~6(b) is the spectrum processed from this image,
which was analyzed in Paper~1.  Like Lyman-$\alpha$, the center of the
chromospheric emission line is highly absorbed by the stellar wind.
In Figure~6(d), cross-dispersion profiles are shown at three different
locations in the line profile.  The first and third of these are near
the flux peaks blueward and redward of the wind absorption.  The
cross-dispersion profiles at both of these locations are narrow
and point-source-like, with widths of about 2.2 pixels.  In contrast,
the second cross-dispersion profile, from the redward edge of the
absorption where wind opacity is signifiant, is roughly twice
as broad, and clearly {\em not} point-source-like.  Note that the
$0.2^{\prime\prime}$ width of the aperture in the cross-dispersion
direction corresponds to about 8 pixels.
The cross-dispersion profiles are fitted with Gaussians, and in
Figure~6(e) the inferred full-width-at-half-maxima (FWHM) are plotted
versus pixel \# across the line profile.  The emission is narrow and
point-source-like at most wavelengths, but not along the red side
of the wind absorption, where the emission is clearly spatially
resolved.  There is naturally a gap in the width measurements where
the wind absorption is saturated and no cross-dispersion width
can be measured.

\begin{figure}[t]
\plotfiddle{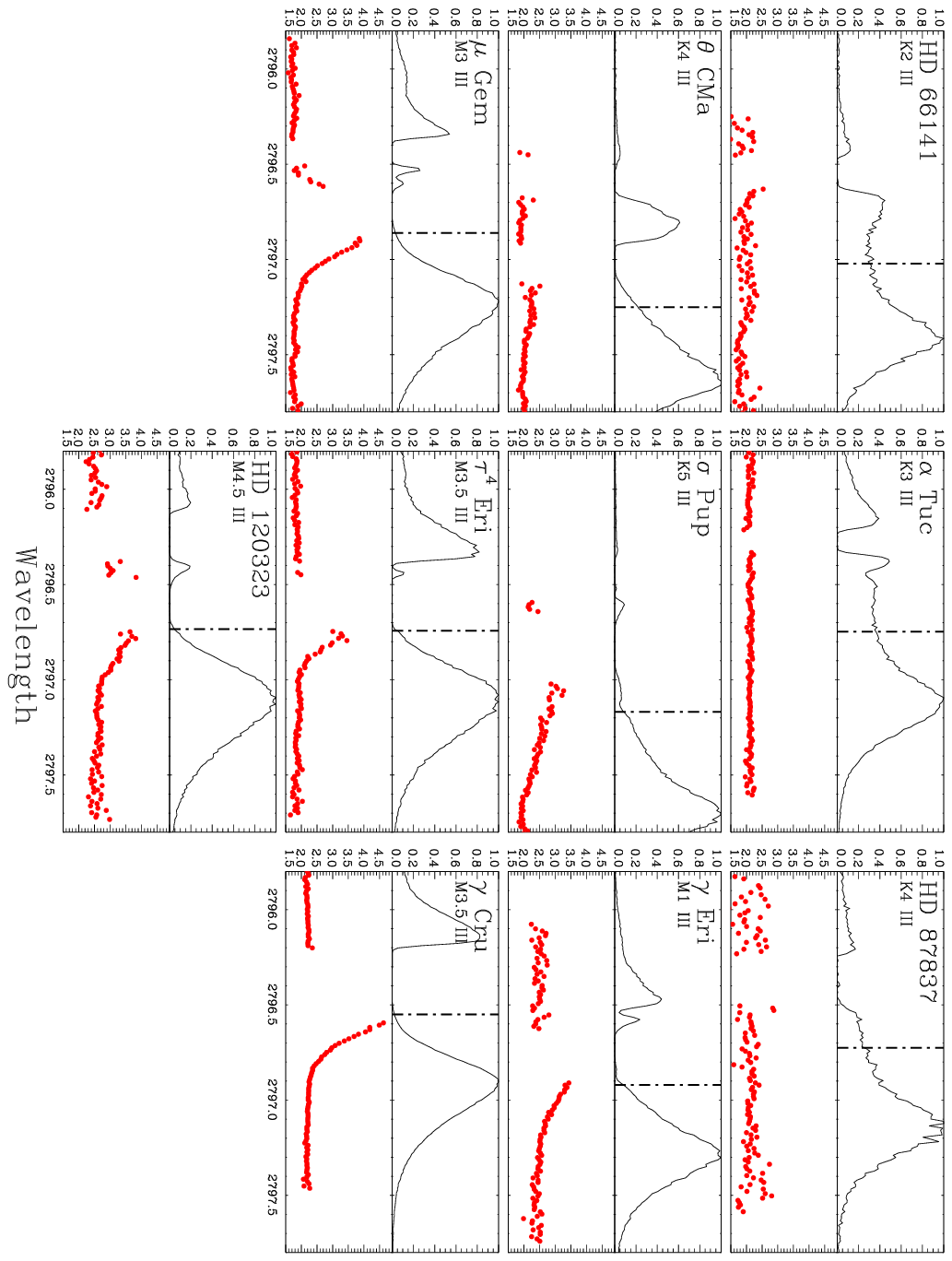}{3.6in}{90}{63}{63}{260}{-55}
\caption{Normalized Mg~II k line spectra (top) and cross-dispersion line
  widths (bottom) are shown for all the red giants in our sample,
  except for $\alpha$~Tau.  The stars are in spectral type order from
  K giants to M giants.  Vertical dot-dashed lines indicate the
  stellar rest frames.  The cross-dispersion widths of the first
  three K giants show no convincing spatial resolution signatures, but
  the other stars show broader widths along the red side of the
  wind absorption, with the signature increasing in strength for
  the M giants.}
\end{figure}
     In Figure~7, we show the Mg~II k line cross-dispersion widths
for all the red giants in our sample, excluding $\alpha$~Tau for
which we have no E230H data.  We note again that these spectra were
discussed at length in Paper~1, with extensive discussion of
what features are wind absorption and what is ISM absorption.
The stars in Figure~7 are in order
of spectral type.  The first three K giants shown (HD~66141,
$\alpha$~Tuc, HD~87837) show no convincing evidence of spatial
resolution at all, with the cross-dispersion widths being narrow
throughout the profile.  A hint of the spatial resolution signature
along the red side of the absorption is seen for $\theta$~CMa, and
this signature becomes more prominent for the later type stars.
For $\mu$~Gem and HD~120323, signatures of spatial resolution are
also seen on the blue side of the absorption, near
2796.5~\AA.  It should be noted that the presence of ISM absorption
is probably preventing similar blue-side spatial resolution
signatures from being seen for
$\tau^4$~Eri and $\gamma$~Cru (see Paper~1).

     The spatial resolution signatures are clearly stronger for the
M giants than the K giants.  There are two obvious reasons for this.
One is that the M giants have stronger, more opaque winds
(see Section~2).  The other reason is that M giant radii are larger
than K giant radii to begin with, and their angular diameters are
therefore generally larger as well (see Table~1).  The star with the
largest cross-dispersion widths is $\gamma$~Cru, which is the nearest
M giant in our sample.

     There are many chromospheric emission lines in the E230H
data that also show wind absorption, for which we can
look for the same kind of spatial resolution signatures that
we see for Mg~II.  For example, the Mg~I line at a vacuum rest
wavelength of 2852.964~\AA\ (all quoted NUV wavelengths below
will also be vacuum wavelengths) shows
wind absorption similar to that of Mg~II, and we find that it has
similar spatial resolution signatures as well.  This is important
evidence that Mg is partially neutral throughout the wind
acceleration region.

\subsection{The NUV Fe~II Lines}

\begin{figure}[t]
\plotfiddle{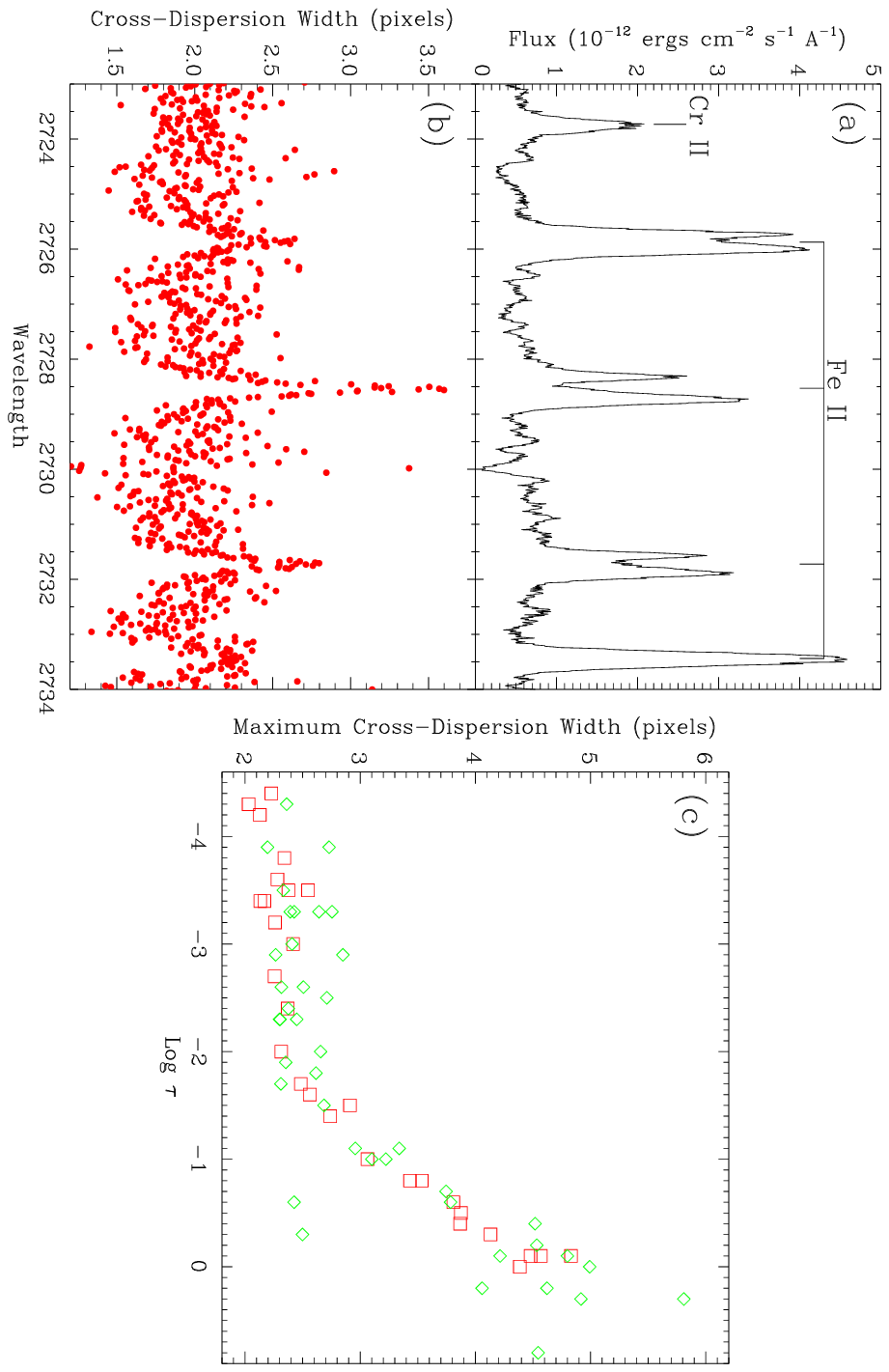}{3.6in}{90}{63}{63}{260}{-55}
\caption{(a) STIS/E230H spectrum of the 2723--2734~\AA\ spectral region
  of $\gamma$~Cru, showing four Fe~II lines with wind absorption of
  varying strength, and one Cr~II line.  (b) Cross-dispersion widths
  (in pixels) plotted across the spectral range shown in (a), with
  spatial resolution signatures apparent for all four Fe~II lines,
  with the broader cross-dispersion signatures seen for the lines
  with stronger wind absorption.  (c) Maximum Fe~II cross-dispersion
  width versus relative line opacity, with more opaque lines clearly
  being more spatially resolved.  Red squares (green diamonds) are
  for lines measured from HST observation ID\#OBKK82020 (\#OBKK81020),
  covering $2620-2888$~\AA\ ($2378-2651$~\AA).}
\end{figure}
     However, it is Fe~II that is the atomic species with by far the
most numerous lines in the E230H spectra, many showing wind
absorption \citep{gr18,ken23}.  In Figure~8(a),
we show the 2723--2734~\AA\ spectral region of $\gamma$~Cru, with
four Fe~II lines and one Cr~II line.  Below this spectrum in
Figure~8(b) we present the cross-dispersion FWHM measurements
for this spectral region.  Peaks are seen for all four Fe~II lines,
with broader widths seen for the lines with stronger wind absorption.

     The Fe~II 2728.3~\AA\ line has the strongest wind absorption
signature, and also the strongest spatial resolution signature with
the cross-dispersion width reaching $\sim 3.7$~pixels.  In contrast,
the Fe~II 2733.3~\AA\ line does not have a clear wind absorption
signature, and has the lowest cross-dispersion width of
only $\sim 2.3$~pixels.  However, even this
line seems to have a cross-dispersion width slightly above that of
a point source.  Note that the Cr~II 2723.6~\AA\ line in this
spectral region shows no spatial resolution signature at all.
Another point to make about the Fe~II lines is that they show the
strongest spatial resolution signatures where the wind absorption
is deepest.  This is not clear for the strongest chromospheric
lines like Mg~II, where the wind absorption is fully saturated
where it is strongest, making it impossible to even look at a
cross-dispersion profile at those wavelengths.

     We have measured cross-dispersion widths for two separate
E230H spectra of $\gamma$~Cru (Obs.\ IDs \#OBKK81020 and \#OBKK82020),
covering a combined wavelength range of $2378-2888$~\AA.  There are
75 Fe~II lines in this region, based on the line list of
\citet{kgc18}.  \citet{kgc18} also provide
relative LTE line opacity values for these lines, assuming
$T_{exc}=6000$~K, and in Figure~8(c)
we plot maximum cross-dispersion widths versus these line opacities.
There is consistent, clear dependence of cross-dispersion width with
relative line opacity, with the cross-dispersion widths increasing
dramatically for $\log \tau>-2$.
%29+50-4

\subsection{The H~I Lyman-$\alpha$ Line}
\begin{figure}[t]
\plotfiddle{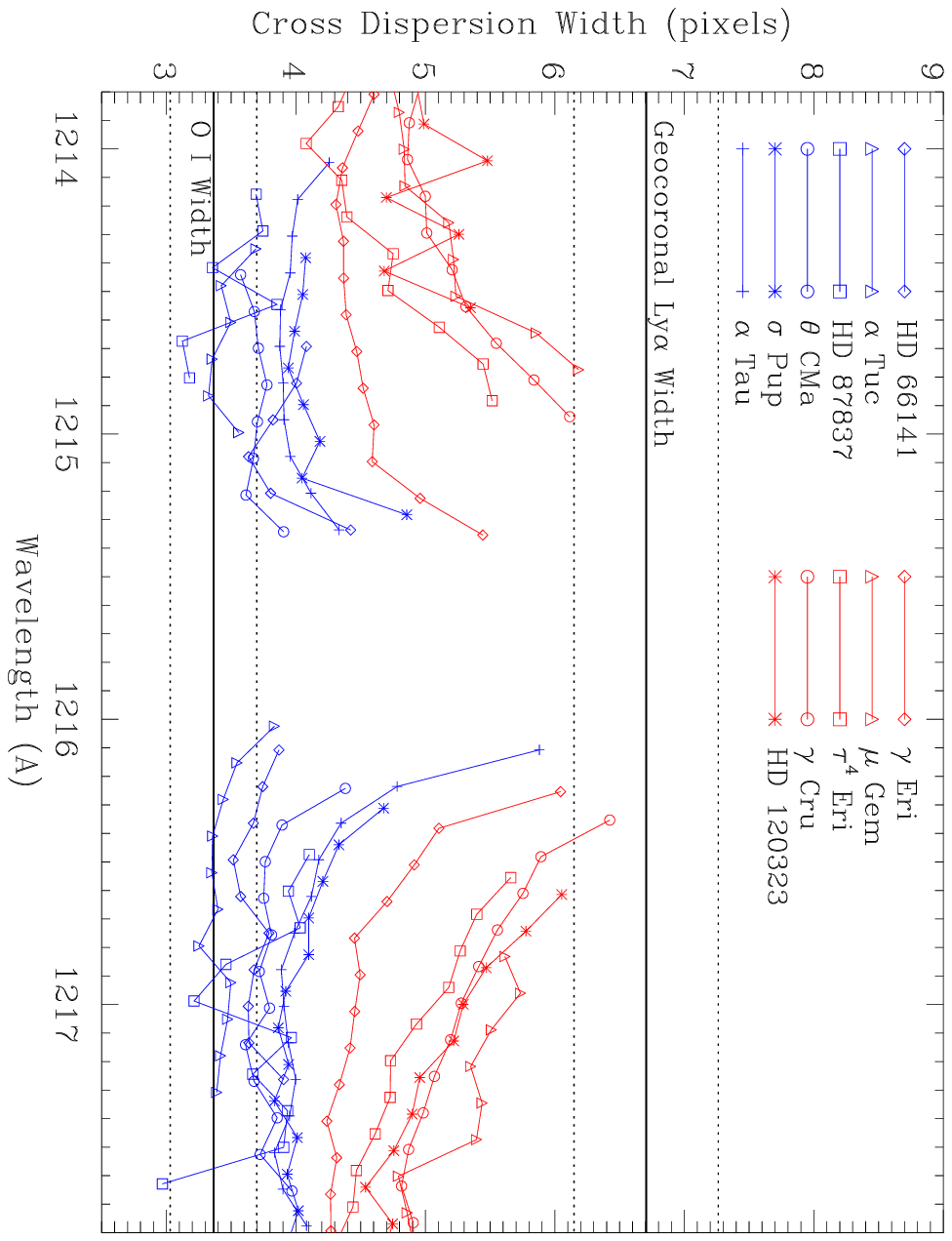}{3.6in}{90}{63}{63}{240}{-55}
\caption{Cross-dispersion widths are measured across the H~I Lyman-$\alpha$
  lines of 11 red giants, with the K giants shown in blue and the M giants
  in red.  The gap in the middle indicates where the wind and/or ISM
  absorption is saturated (see Figure~1).  The widths are also compared
  with cross-dispersion widths measured for geocoronal emission seen in
  these spectra, which uniformly fills the aperture
  ($FWHM=6.7\pm 0.6$~pixels), and also a presumed point-source
  cross-dispersion width measured at the peak flux of the
  O~I $\lambda$1306 line ($FWHM=3.4\pm 0.3$~pixels).  The stellar
  cross-dispersion widths lie comfortably in between the two expected
  extremes, with the M giant emission being more spatially resolved than
  the K giants.}
\end{figure}
     Returning to Lyman-$\alpha$, cross-dispersion widths are measured
across the broad H~I Lyman-$\alpha$ lines in the E140M echelle spectra,
and in Figure~9 these widths are shown for all the stars in our sample.
The wavelength gap near line center is naturally
indicative of where the wind and/or ISM absorption is saturated and
no flux is therefore available to allow for a cross-dispersion width
measurement (see Figure~1).  These stellar emission widths are compared
with cross-dispersion widths of geocoronal Lyman-$\alpha$ emission,
which is visible in Figure~1, and with cross-dispersion
widths measured at the peak of the O~I $\lambda$1306 line, where we
believe the emission is point-source-like.  (In the next subsection,
we will show that O~I is spatially resolved within the wind absorption
region, but not where the line has maximum flux.)

     For the geocoronal emission, which should be uniformly filling
the $0.2^{\prime\prime}\times 0.2^{\prime\prime}$ aperture, we find
a mean cross-dispersion width and standard deviation of
$FWHM=6.7\pm 0.6$~pixels for the 11 stars in our sample.
For the O~I lines, at peak flux we find a presumed point source width
of $FWHM=3.4\pm 0.3$~pixels.  The stellar Lyman-$\alpha$ cross-dispersion
widths in Figure~9 fall comfortably in between these two extremes.
As is the case for Mg~II in Figure~7, the M giants have significantly
broader cross-dispersion profiles in Lyman-$\alpha$ than the K giants.
At their broadest, some of the M giant widths approach that of the
geocorona.  The spectral range over which the emission is spatially
resolved is much broader than Mg~II, particularly for the M giants,
with the emission being spatially resolved far from line center on
both sides of the line.  This is not surprising considering that
the stellar wind opacity will be orders of magnitude higher for
H~I Lyman-$\alpha$ than it is for Mg~II.  Although less spatially
resolved than the M giants, the K giants all show at least a hint of
spatial resolution, with widths at least slightly broader than the
point-source width at some wavelengths.

\subsection{Other FUV Lines}
\begin{figure}[t]
\plotfiddle{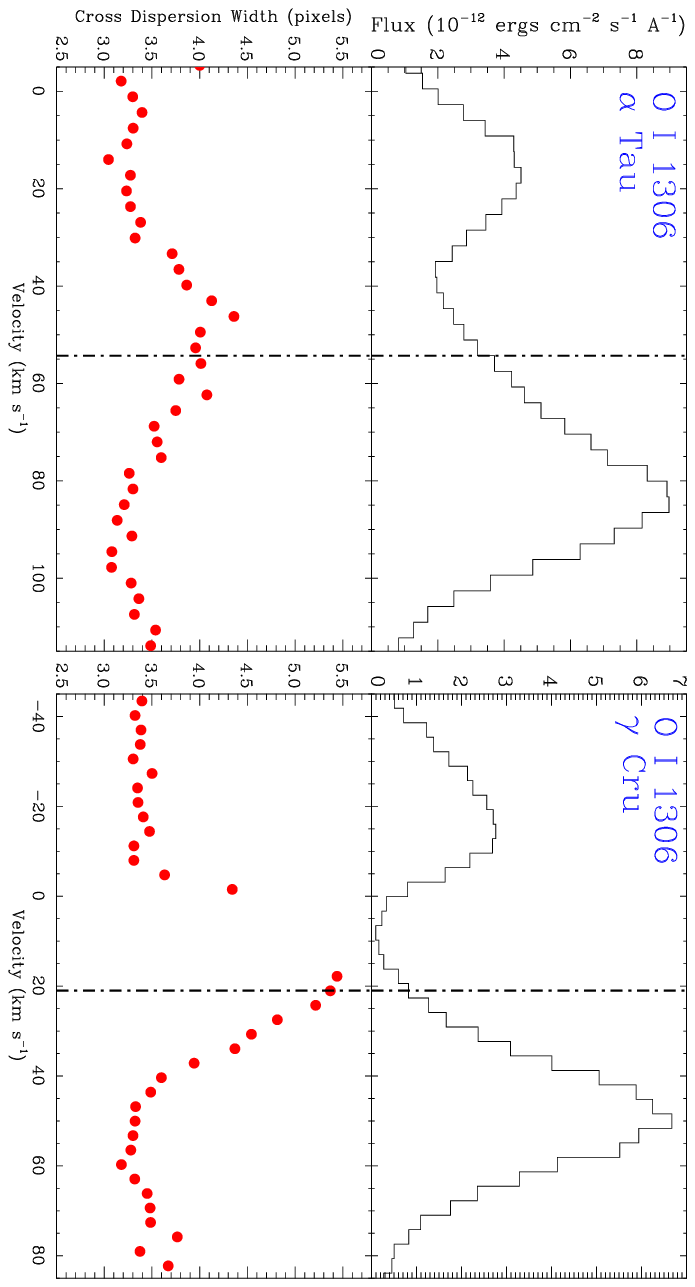}{3.0in}{90}{63}{63}{240}{-75}
\caption{STIS/E140M spectra of the O~I $\lambda$1306 line are shown
  for $\alpha$~Tau (K5~III) and $\gamma$~Cru (M3.5~III), plotted on
  a heliocentric velocity scale, with vertical dot-dashed lines
  indicating the stellar rest frame.  Below the spectra are
  cross-dispersion line widths, showing evidence for spatially
  resolved emission, particularly for $\gamma$~Cru.}
\end{figure}
     The 1140--1735~\AA\ spectral range of the E140M spectra contains
other lines with wind absorption that are worthy of attention.  By far
the strongest of these are the O~I triplet lines, with rest wavelengths
of 1302.2~\AA, 1304.9~\AA, and 1306.0~\AA.  The wind absorption
features in these lines are essentially identical, although ISM
absorption can complicate the ground-state $\lambda$1302 line profile.
The cross-dispersion width behavior also appears similar for all three
lines, and resembles the pattern seen for Mg~II in Figure~7.
In Figure~10, we explicitly show
the O~I $\lambda$1306 line profiles and cross-dispersion width
measurements for $\alpha$~Tau and $\gamma$~Cru, the brightest K and M
giants in our sample, respectively.

     The highest temperature lines with wind absorption are the
C~II $\lambda\lambda$1335, 1336 lines.  However, for most of our
spectra these lines are too noisy to measure cross-dispersion
widths.  The exception is $\gamma$~Cru, and for those data we can
confirm cross-dispersion width behavior similar to that of the
O~I line in Figure~10.

     The HST constraints on velocity-dependent spatial extent for many
NUV and FUV line hold significant promise for testing red giant wind
models.  A full study of this nature is outside the scope of this paper,
but we can offer a sample of such a data/model comparison using a
wind model that has recently been constructed for $\gamma$~Cru.  The model,
which considers not only various UV line profiles but also radio data,
will be published in a companion paper (Harper et al.\ 2024,
in preparation).  Spherically symmetric radiative transfer models
have been computed for wind-scattered line profiles of the O~I resonance
triplet lines at 1302, 1304, 1306\AA, the Mg~II h line, Mg~I at 2851~\AA,
and a sample of NUV Fe~II emission lines.
The line profiles are used to constrain each
ion's mass-loss rate, and the wind's radial velocity and turbulence profiles.
This modeling is fully analogous to the analysis of $\alpha$~Boo's wind
described by \citet{gmh22}.  Radio continuum data are used to
constrain $\gamma$~Cru's wind's total mass-loss rate and temperature. 

\begin{figure}[t]
\plotfiddle{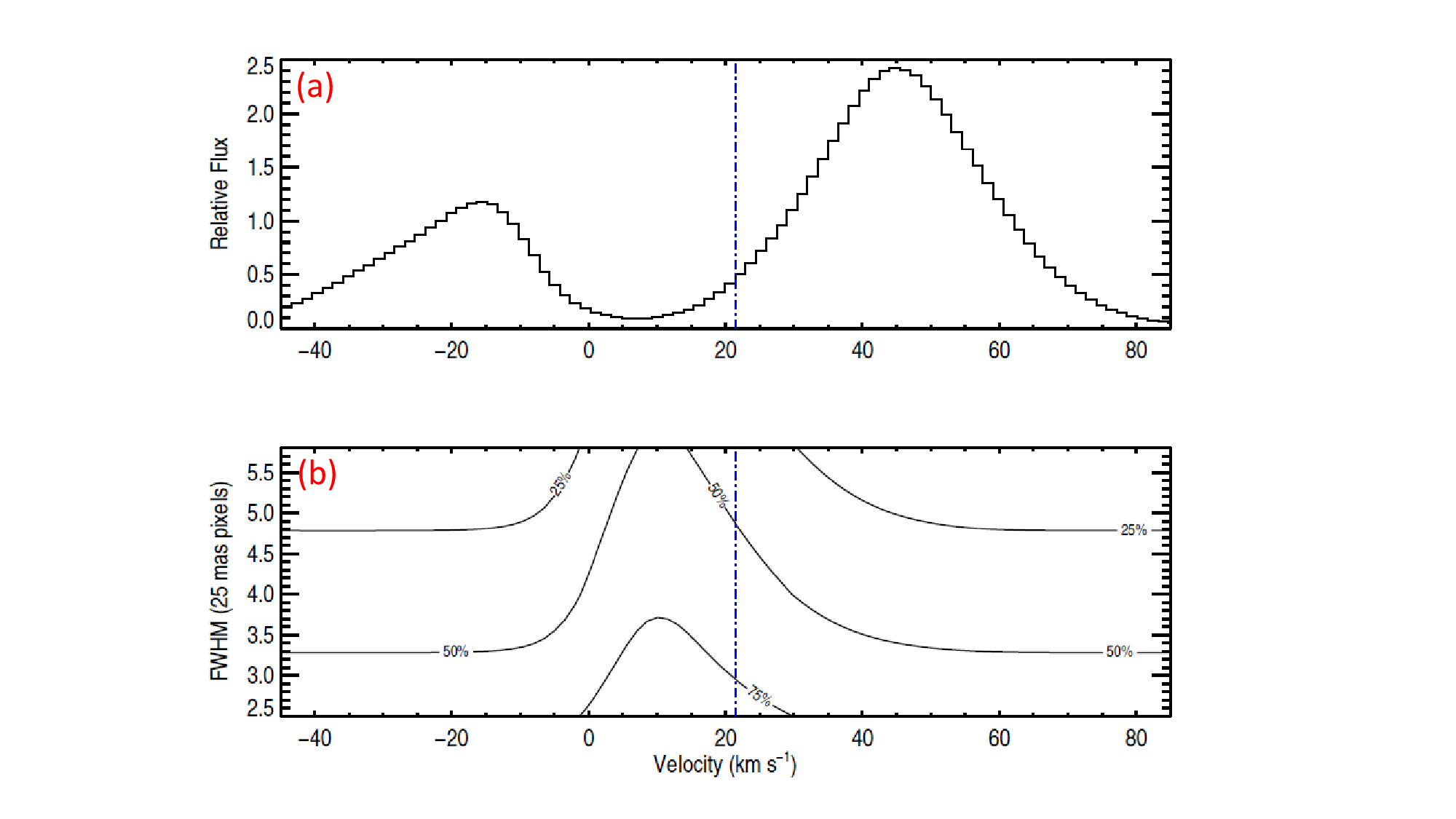}{3.3in}{0}{55}{55}{-260}{-20}
\caption{(a) O~I 1306~\AA\ line profile in a stellar rest
  frame predicted by a radiative transfer wind model for $\gamma$~Cru,
  which agrees well with the observed profile shown in Figure~10.
  (b) Spatial extent contours for the O~I emission as a function of
  heliocentric velocity predicted by the model.  The vertical blue
  line indicates the stellar rest frame.  The FWHM
  represented by the 50\% contour can be compared
  with the observed FWMH values in Figure~10, showing good agreement.}
\end{figure}
     We here use these radiative transfer calculations to
predict how the wind scattered emission for each line appears on the
sky, or through the HST-STIS aperture.  In Figure~11, we present an example
for the O~I 1306\AA\ line.  Figure~11(a) shows the model O~I 1306~\AA\ line
profile, which agrees with the $\gamma$~Cru line profile in Figure~10.
The calculated wind emission has been convolved
with a Gaussian to mimic the HST PSF as reflected by the spatial dispersion
seen in the nearby stellar continuum, and then binned into STIS's 25~mas
pixels and summed in the spectral dispersion direction.  Figure~11(b) shows
contours of the spatial extent of the emission as a function of velocity.

     At velocities of $V<-10$ km~s$^{-1}$ and $V>40$ km~s$^{-1}$, no
significant spatial extent is expected, consistent with the observations
in Figure~11.  The 1.6 pixel half-width in these regions corresponds with
a $FWMH=3.2$ pixels, consistent with the point source widths observed and
shown in Figures 9-10.  The peak width observable in the data is near
$V=20$ km~s$^{-1}$, where $FWHM=5.4$ pixels (see Figure~10), in
good agreement with the model
predictions near $V=20$ km~s$^{-1}$ in Figure~11(b).
Extended regions of the wind emission are not captured by the
STIS $0.2^{\prime\prime}\times 0.2^{\prime\prime}$ aperture, but within the
aperture the agreement is excellent.  The wavelength dependence of the
empirical spatial extents of wind scattered lines could be used as
additional constraints in future wind modeling of nearby red giants.

\section{HST Long-Slit H~I Lyman-$\alpha$ Observations of
  $\gamma$~Cru and $\alpha$~Tau}

     In the previous section, we showed how STIS echelle spectra
provide a surprising amount of information about the spatial extent
of chromospheric emission from red giant winds, despite not being
intended to provide such information.  However, these data have
obvious limitations.  With an aperture only
$0.2^{\prime\prime}$ wide in the spatial direction, the echelle data
cannot be used to measure the full extent of the emission around the
stars, or to study how uniform the emission is around the stars.
Such information can only be provided by imaging or long-slit
spectroscopy.

     In order to provide more detailed information about the spatial
extent of red giant wind emission in the UV, we have used HST to
obtain long-slit Lyman-$\alpha$ spectra for $\alpha$~Tau and
$\gamma$~Cru.  These two targets are meant to be representative of
K and M giants, respectively, and were chosen mostly due to their
brightness and close proximity, allowing us to search for spatially
extended emission as far from the stars as possible.  Both
observations were made over four HST orbits with the G140M grating,
using the $52^{\prime\prime}\times 0.2^{\prime\prime}$ long slit,
covering a wavelength region of $1194-1249$~\AA.
The 10,578~s $\alpha$~Tau observations were completed on
2021~January~1, and the 11,510~s $\gamma$~Cru data were
acquired on 2021~January~8.

\begin{figure}[t]
\plotfiddle{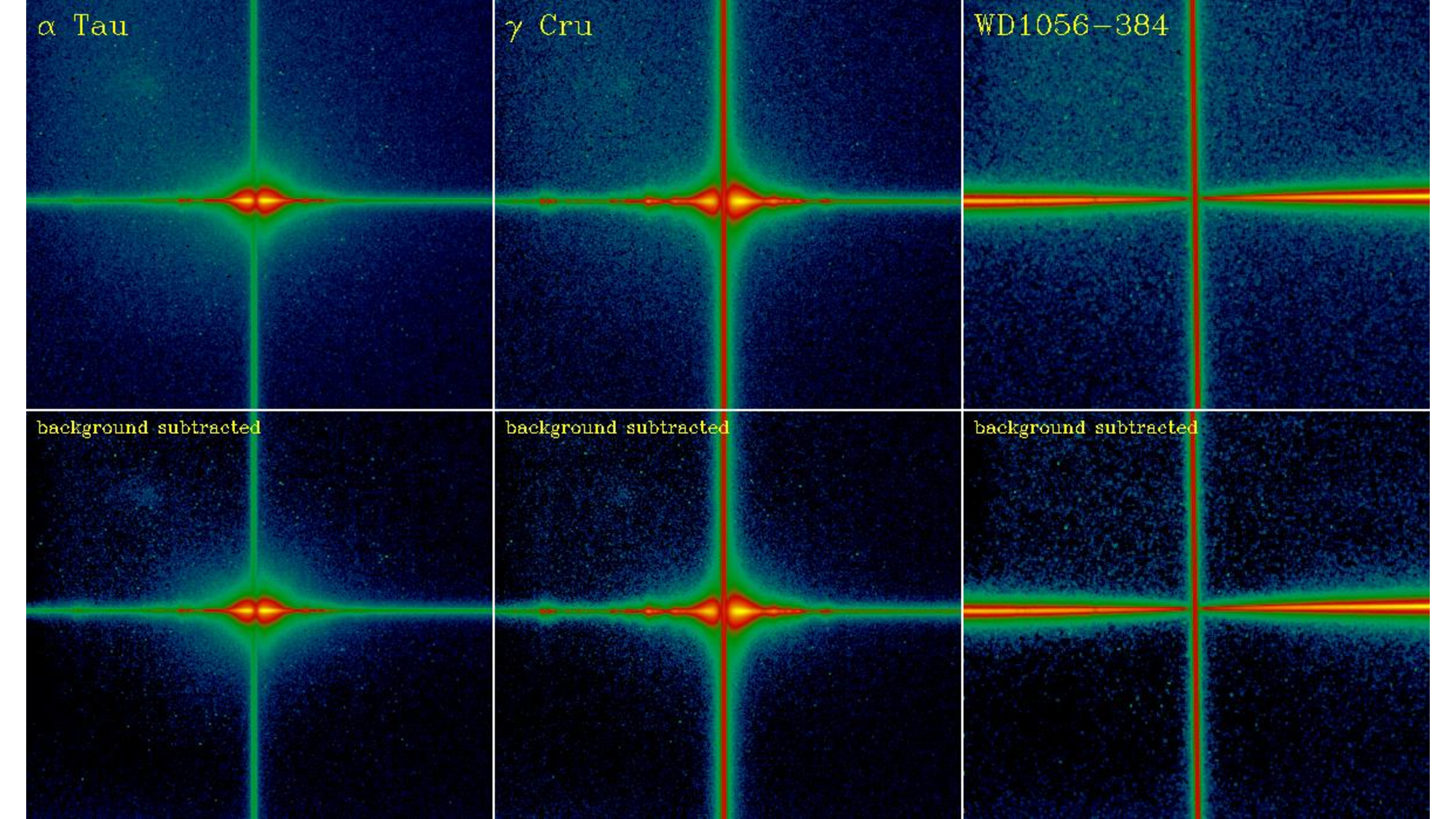}{3.4in}{0}{50}{50}{-240}{-10}
\caption{HST/STIS long-slit spectral images of the H~I Lyman-$\alpha$
  line, using the G140M grating.  Data are shown for the two
  red giants of interest, $\alpha$~Tau and $\gamma$~Cru, as well as
  a white dwarf target, WD1056-384, which is used to establish
  the instrumental PSF.  The images are shown before and
  after background subtraction.  The vertical stripe is geocoronal
  Lyman-$\alpha$ emission.}
\end{figure}
     The HST/STIS images showing the G140M long-slit data are shown
in Figure~12.  The horizontal spectral stripe is clearly visible
for both $\alpha$~Tau and $\gamma$~Cru, dominated by the broad,
bright H~I Lyman-$\alpha$ emission near image center, with
the center of the profile absorbed by the stellar wind and/or ISM.
In the middle of this absorption lies a vertical stripe of
emission, which is geocoronal emission filling the entire long-slit
aperture.

     In Figure~12, the $\alpha$~Tau and $\gamma$~Cru data are
compared with a STIS/G140M Lyman-$\alpha$ observation of a white
dwarf (WD), WD1056-384, taken from the HST archives.  In order to
properly assess the spatial extent to which emission can be
observed from our red giants, we need to know the PSF
of the STIS/G140M data well.  The WD
data are used here to provide this PSF information.  The
WD1056-384 spectrum shown in Figure~12 is that of continuum
emission from the hot WD, with very broad
stellar Lyman-$\alpha$ absorption.

     In the process of exploring the spatial extent of the
Lyman-$\alpha$ emission it became clear that it was necessary to
address the uneven scattered light pattern that underlies all three
of the STIS/G140M observations in Figure~12, which is brightest
in the upper left quadrant.  For each image, we modeled this
background by first measuring it along 50-pixel wide
summed vertical swathes, and then fitting this emission profile with
second-order polynomials.  This analysis avoided regions within
150 pixels of the horizontal spectral stripe and the vertical
geocoronal emission stripe, but the background model could be
interpolated underneath the ignored regions, allowing for the
necessary scattered light correction in the region of interest near
the actual spectra.  The resulting images after background subtraction
are also shown in Figure~12.  The noise left behind by the scattered
light is still apparent, particularly in the upper left quadrant.
The red giant data also show a small spot of emission in that quadrant
that is not removed, but is far enough from the spectrum to
be unimportant.  This is presumably a ghost of the central, bright
Lyman-$\alpha$ emission source at image-center.

\begin{figure}[t]
\plotfiddle{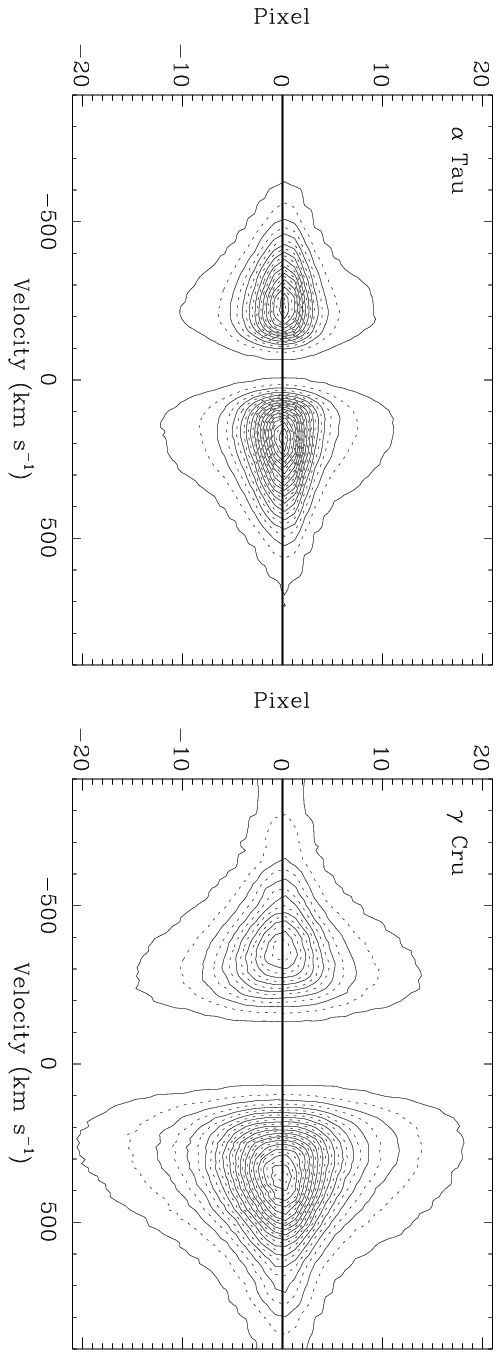}{1.9in}{90}{63}{63}{240}{-175}
\caption{Contour plots of the $\alpha$~Tau and $\gamma$~Cru
  long-slit H~I Lyman-$\alpha$ spectral images from Figure~12,
  plotted on a velocity scale in the stellar rest frames.
  Stellar wind and ISM are responsible for the central
  absorption.  The 30 contours are evenly spaced between the
  flux maximum and a floor defined at 3\% of that maximum.
  The emission of $\gamma$~Cru is
  clearly more extended than that of $\alpha$~Tau.}
\end{figure}
     In Figure~13, the G140M Lyman-$\alpha$ emission is displayed
in a contour plot.  In making these plots, the emission is first
normalized relative to the emission peak, and pixels with fluxes
lower than 3\% of this peak are ignored in order to focus attention
on the higher S/N part of the line.  This also
effectively removes the geocoronal emission.  The broader spatial
extent of $\gamma$~Cru relative to $\alpha$~Tau is easily apparent,
consistent with the inferences from the echelle data in Section~3
(see Figure~9).  The $0.2^{\prime\prime}$ aperture used for the
echelle data is about 8 pixels wide.  Although some of the red giant
emission falls outside this aperture, most of the Lyman-$\alpha$
flux falls within it even for $\gamma$~Cru.  This means that although
the Lyman-$\alpha$ fluxes reported in Table~1 will be underestimates
to some extent due to the spatial resolution of the emission, this
will only be a $30$\% effect at most.

     Long-slit spectroscopy only provides spatial information along
one direction, along the slit.  The Lyman-$\alpha$ emission in
Figure~13 is relatively symmetric about the star along that direction.
For $\gamma$~Cru there is a hint of the emission being slightly
brighter below the zero pixel line, but by no more than $\sim 10$\%.
With a plate scale of 0.0246$^{\prime\prime}$ per pixel, one pixel
width conveniently corresponds to roughly one stellar diameter for
both $\alpha$~Tau and $\gamma$~Cru, based on the stellar radii and
distances in Table~1.  Figure~13 shows emission extending out to at
least 20 pixel from the star for $\gamma$~Cru, corresponding to about
40~R$_*$.

\begin{figure}[t]
\plotfiddle{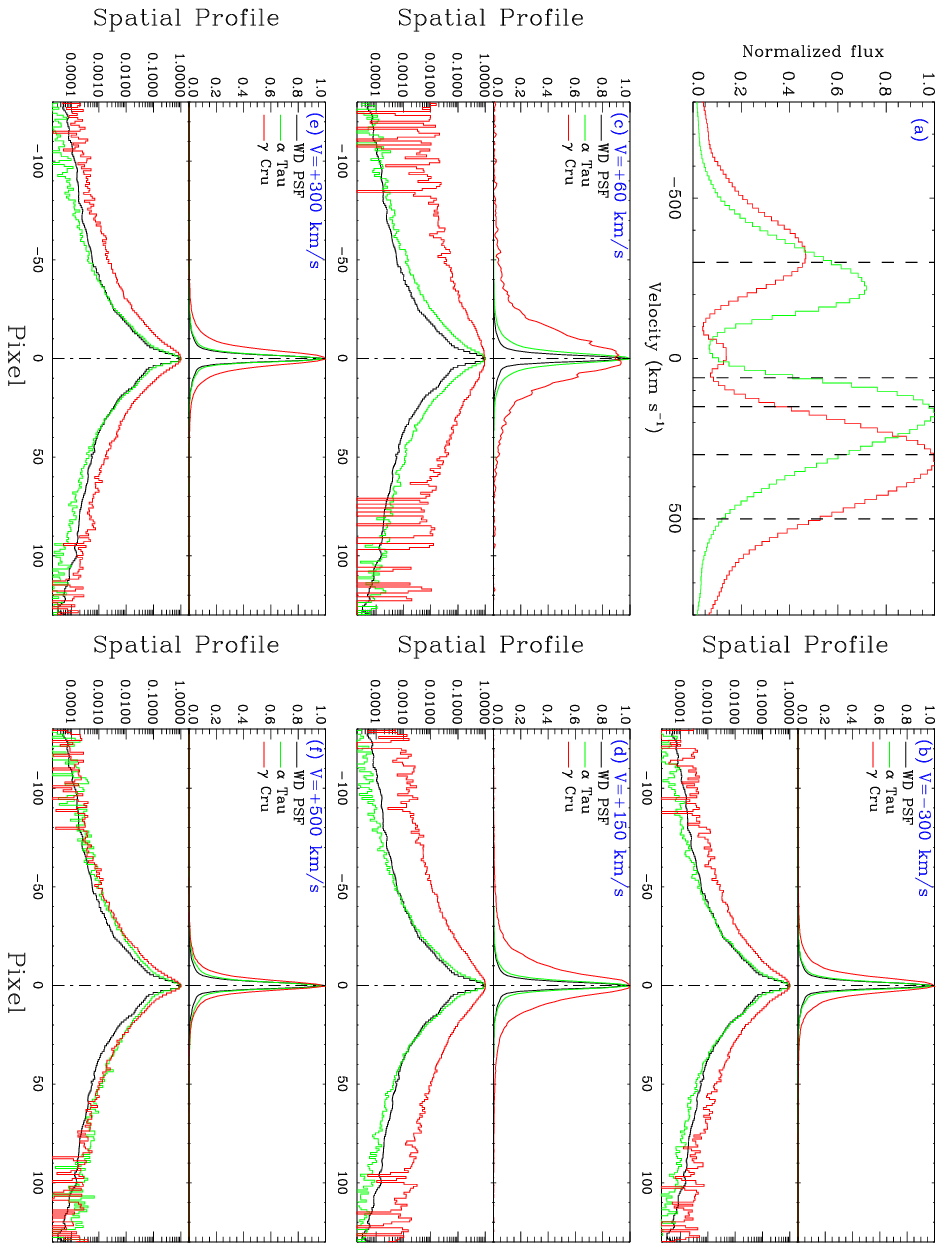}{3.8in}{90}{70}{70}{280}{-60}
\caption{(a) Normalized spectra of $\alpha$~Tau (green) and
  $\gamma$~Cru (red), extracted from the long-slit G140M observations
  shown in Figure~12 using a broad spatial window, plotted on
  a velocity scale in the stellar rest frame.  The peak
  near 0 km~s$^{-1}$ for $\gamma$~Cru is geocoronal emission.
  Vertical dashed lines indicate velocities where spatial
  emission profiles are shown in the following panels.
  (b-f) Each panel shows spatial emission profiles at one
  of the five velocities indicated in (a), for both $\alpha$~Tau
  and $\gamma$~Cru, using both a linear (top) and a logarithmic
  (bottom) scale.  These are compared with the PSF inferred from
  the WD1056-384 data (see Figure~12).  }
\end{figure}
     Figure~14 provides more detailed intensity tracings along the
spatial direction to explore just how far the emission extends.
Figure~14(a) first shows normalized spectra extracted using a broad
window along the spatial direction, for both $\alpha$~Tau and
$\gamma$~Cru.  The geocoronal emission peak is clearly seen for
$\gamma$~Cru near 0 km~s$^{-1}$.  Geocoronal emission is present
for $\alpha$~Tau as well, but is much weaker relative to the
stellar fluxes and is therefore less apparent.  These Lyman-$\alpha$
profiles can be compared with the STIS/E140M spectra in Figure~1.
Note that the G140M spectra have significantly lower spectral
resolution than E140M.

     Vertical dashed lines in Figure~14(a) indicate the five
velocities where subsequent panels explore the spatial
extent of the emission.
%In each case, we sum up fluxes within
%$\pm 25$ km~s$^{-1}$ of the central velocity, in order to
%increase S/N.
We assume a flat background underneath the emission,
which is measured using pixels $>200$ pixels from the star.
This is necessary despite the background subtraction described
above and shown in Figure~12, because of the presence of the
vertical geocoronal emission stripe, which has wings that extend
from the stripe in the wavelength direction, creating a
relatively flat background that must be removed.

     In Figure~14(b-f), the resulting intensity tracings are
shown for $\alpha$~Tau, $\gamma$~Cru, and the WD PSF
comparison star, WD1056-384.  For the WD, we compute a single
cross-dispersion PSF to represent the PSF at all velocities, though
we consider only the right half of the spectral range shown
in Figure~12, avoiding the left half where the scattered light
background is highest.  Both linear and logarithmic scales
are shown, with the latter being necessary to reveal behavior
in the far wings of the spatial profiles.
%The profiles are
%shown on a pixel scale, but vertical dotted lines indicate
%how this translates to distances from the star in stellar radii,
%with the scale shown to the left for $\alpha$~Tau and to the
%right for $\gamma$~Cru.  As noted above, the scales are actually
%very similar since 1 pixel is about 2~R$_*$ in both cases.

     We focus first on the $\alpha$~Tau data.
At $V=-300$ km~s$^{-1}$ and V=$+300$ km~s$^{-1}$,
the $\alpha$~Tau profile agrees fairly well with the
WD profile, except in the far wings ($>50$~pixels from center)
where the WD PSF appears stronger.  Uncertainties in the
background subtraction will be a factor this far into PSF
wings.  There is evidence that
$\alpha$~Tau is slightly broader than
the WD near the star, suggesting some spatial extent to the
emission.  This is consistent with the evidence for modest
spatial extent of the STIS/E140M emission far from line center
in Figure~9.  Stronger evidence of
spatial extent to the emission is seen at $V=+150$ km~s$^{-1}$.
However, even here the spatial extent of the emission seems to be
within $\pm 30$~pixels.  With $\sim 2$ R$_*$ per pixel, this
corresponds to $\pm 60$ R$_*$.  The only velocity where $\alpha$~Tau
emission may be detected $>30$~pixels from the star is at
$V=+60$ km~s$^{-1}$.  This behavior is consistent with the
dramatic rise in cross-dispersion width seen for the E140M data in
Figure~9 near the red side of the Lyman-$\alpha$ absorption, though
the E140M data is really only measuring the dramatic increase in
spatial extent close to the star.  At $V=+60$ km~s$^{-1}$, the long
slit data suggest that detectable emission extends well beyond the edge
of the $0.2^{\prime\prime}$ aperture used in the E140M observations.

     At first glance, the $V=+500$ km~s$^{-1}$ $\alpha$~Tau profile
would seem to surprisingly imply very extended emission in the red wing
of the $\alpha$~Tau Lyman-$\alpha$ profile, different from what is
observed at V=$+300$ km~s$^{-1}$.  However, in this case we believe
that this is illusory, with the extra emission in the profile wings
being not from spatially extended emission, but from the wings of the
PSF in the {\em spectral} direction, extending redward from the
brightest part of the line.  This illustrates the danger in trying to
infer spatial extent in low-flux parts of a spectral line that could
end up significantly contaminated by the PSF wings of the brighter parts
of the line profile.  It is safer if this kind of analysis is limited
to parts of the spectral line profile with at least a third of the
peak line flux.

     As expected, the Lyman-$\alpha$ emission is more extended
for $\gamma$~Cru than for $\alpha$~Tau.  The spatial extent is beyond
that of both the WD and $\alpha$~Tau at all velocities, and at all
distances from the star.  It should be noted that the
$V=+60$ km~s$^{-1}$ $\gamma$~Cru profile should be regarded with
suspicion for the same reasons as the $V=+500$ km~s$^{-1}$ $\alpha$~Tau
profile.  Not only is the line flux low here relative to the nearby
peak of the Lyman-$\alpha$ line, but the background is very high due
to close proximity to the geocoronal emission stripe.  Nevertheless,
at $V=+150$ km~s$^{-1}$, the $\gamma$~Cru emission is clearly well
above the PSF (and $\alpha$~Tau) out to at least $\sim 100$~pixels
(e.g., $\sim 200$~R$_*$), at which
point the emission is becoming too noisy to distinguish from the
background.  As was the case for $\alpha$~Tau, the extent of the
emission in the long-slit data mirrors what might be inferred from
the behavior of the cross-dispersion widths measured from the the
E140M data in Figure~9, with the greatest spatial extent seen near
the red side of the absorption.

\begin{figure}[t]
\plotfiddle{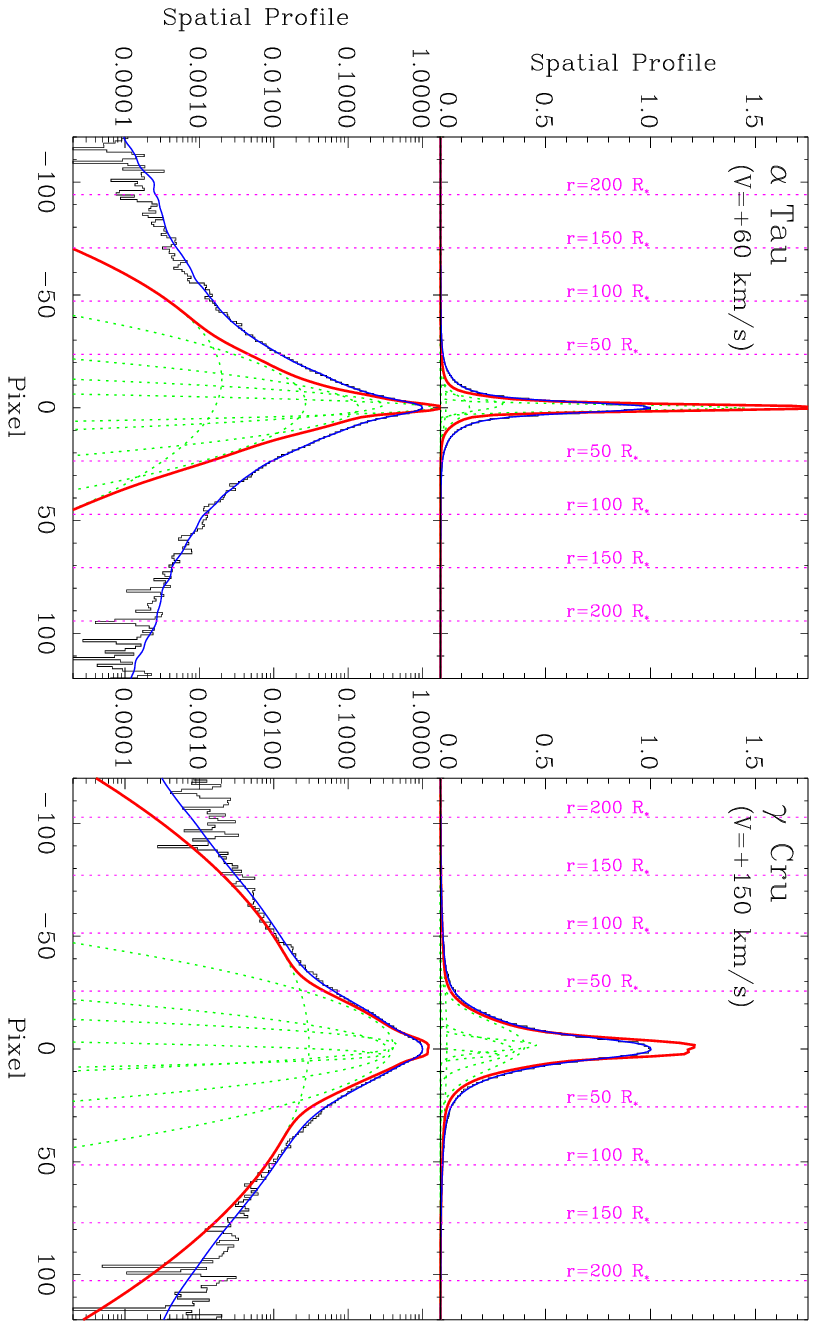}{3.2in}{90}{65}{65}{250}{-70}
\caption{The $V=+60$ km~s$^{-1}$ $\alpha$~Tau and $V=+150$ km~s$^{-1}$
  $\gamma$~Cru spatial profiles from Figure~14 are shown using both
  linear and logarithmic scales.  As a method of deconvolving the
  profiles with the spatial PSF, a five-Gaussian fit is made to
  both profiles, with the sum of the Gaussians convolved with the
  WD PSF from Figure~14 before comparing with the data.  The
  resulting Gaussians and their sum are shown as dotted green
  and solid red lines, respectively.  Convolving the sum with
  PSF yields the blue lines that fit the data.  The red lines
  represent our estimates of deconvolved spatial profiles.
  Vertical dotted lines indicate the distance scale in stellar radii.}
\end{figure}
     The spatial profiles with the strongest and most believable
broad spatial extents are the $V=+60$ km~s$^{-1}$ $\alpha$~Tau
profile and the $V=+150$ km~s$^{-1}$ $\gamma$~Cru profile.  For
these two profiles we attempt a true deconvolution
assuming the WD PSF.  We use a forward modeling approach, where
we simply perform a five-Gaussian fit to the spatial profiles,
with four Gaussians with fixed FWHM widths of 3, 6, 12, and
24 pixels; and the final fifth Gaussian allowed to have any width.
The sum of the five Gaussians is convolved with the WD PSF from
Figure~14 before comparing with the data.  The resulting fits
are shown in Figure~15.  We attach no meaning to the individual
Gaussians, but their sum (red lines in the figure)
represents an estimate of the deconvolved spatial profiles of
the Lyman-$\alpha$ emission at these particular velocities
within the spectral profile.

     In the wings of the spatial profile, the ratio of the red
and blue lines in Figure~15 represents an estimate of the real
emission relative to that from the PSF.  We arbitrarily define the
extent of believable stellar emisison to be where this ratio
is at least 0.33, i.e.\ where the actual emission is at least half
that of the PSF wing.  For $\gamma$~Cru, this point is at
$r=198$~R$_*$ ($r=188$~R$_*$) on the left (right) side.  Taking
the average leads to an estimate of the extent of observable
Lyman-$\alpha$ emission of $r=193$~R$_*$.

     In contrast, the $\alpha$~Tau deconvolution in Figure~15
implies that most of the observed wing emission is largely
from the PSF, and the emission is much less extended than for
$\gamma$~Cru.  The analysis implies that the left wing is stronger
than the right, which seems in part to be an artifact of an
asymmetry in the WD PSF.  It is unclear if this is a real
characteristic of the PSF or if this is an effect of the
uneven scattered light background seen in Figure~12, with
our efforts to remove it being imperfect.  Regardless, if we
use the same criterion used above for $\gamma$~Cru, we find
the limits of believable stellar emission to be
$r=63$~R$_*$ ($r=26$~R$_*$) on the left (right) side, with
an average of $r=44$~R$_*$.  Note that most of the asymmetry in
the deconvolved spatial profile lies outside the narrow range
where we are more confident in the reality of extended
Lyman-$\alpha$ emission.

%\section{???Modeling the Wind of $\gamma$~Cru???}

\section{Summary}

     We have presented here a new analysis of recent HST
spectra of K2-M4.5~III red giant stars, focusing on observational
diagnostics of the strong chromospheric winds of these stars.
The observations considered include E140M and E230H spectra
from our full sample of 11 stars, and long-slit G140M
spectra of $\alpha$~Tau and $\gamma$~Cru.  Our findings can be
summarized as follows:
\begin{enumerate}
\item The absorption seen in the H~I Lyman-$\alpha$ lines has
  been studied to ascertain whether the absorption is wind
  absorption, ISM absorption, or a combination.  Measurements
  of wind and ISM H~I column densities are then made.  Estimates
  of stellar mass-loss rates, $\dot{M}$, are then made from the
  wind $N_H$ measurements.  The M giants have estimated mass-loss
  rates of $\dot{M}=(14-86)\times 10^{-11}$ M$_{\odot}$~yr$^{-1}$,
  while the K giants with detected wind absorption have weaker
  winds with $\dot{M}=(1.5-2.8)\times 10^{-11}$ M$_{\odot}$~yr$^{-1}$.
\item The HD~87837 line of sight shows a surprisingly high ISM
  column density of $\log N_H=19.53$, presumably due to its
  location behind the Local Leo Cold Cloud.
\item We seem to require a high velocity wind absorption
  component to explain the $\alpha$~Tuc (K3~III) Lyman-$\alpha$
  absorption, perhaps similar to the fast wind component inferred
  for the hybrid chromosphere star $\gamma$~Dra.
\item The Lyman-$\alpha$ analysis also provides measurements of
  chromospheric line fluxes for this important line.  We report
  two such fluxes in Table~1, one being simply the observed
  flux, and one based on a line profile reconstructed assuming
  the observed Lyman-$\alpha$ absorption is pure absorption and
  not just due to wind scattering.  The latter are
  more consistent with coronal giants in Mg~II vs.\ Ly-$\alpha$
  flux-flux plots, which is surprising in that it would imply
  that stellar winds are actually destroying Lyman-$\alpha$
  photons and not just scattering them.
\item We have searched for high temperature C~IV and Si~III
  emission in our FUV spectra.  For $\alpha$~Tau, our new
  G140M data provide a detection of Si~III emission, to add
  to the previously reported C~IV detection.  The only other
  stars with clearly detected C~IV and Si~III are $\sigma$~Pup
  and $\alpha$~Tuc, but these are spectroscopic binaries.
  For $\sigma$~Pup, the central wavelengths of the lines are
  more suggestive of the companion being the source,
  but for $\alpha$~Tuc the situation is ambiguous.
\item Extensive evidence is provided that many of the
  chromospheric lines with wind absorption present in the E230H
  and E140M spectra (e.g., Mg~II h \& k, Mg~I, Fe~II,
  H~I Lyman-$\alpha$, O~I, C~II) are actually spatially resolved,
  at least at some locations in the line profiles, with
  cross-dispersion profiles that are broader than those of a
  point source.  The cross-dispersion signatures of spatial
  resolution are much stronger for the M giants compared with
  the K giants, consistent with the presence of stronger,
  more opaque winds for the M giants.
\item For $\gamma$~Cru, cross-dispersion measurements of an
  extensive set of NUV Fe~II lines demonstrates that
  cross-dispersion width correlates very well with
  wind opacity in the line.
\item For $\gamma$~Cru, a full radiative transfer wind model is
  available, and this model is successful at reproducing both the
  line profile of the O~I 1306~\AA\ line and its
  velocity-dependent spatial extent.
\item The long-slit G140M Lyman-$\alpha$ spectra of $\alpha$~Tau
  and $\gamma$~Cru provide more detailed information about
  the spatial extent of the wind emission signatures than
  the narrow-slit E140M/E230H data can provide.  We estimate
  limits for the extent of detectable emission, which are
  $r=193$~R$_*$ for $\gamma$~Cru and $r=44$~R$_*$ for $\alpha$~Tau.
\end{enumerate}

\acknowledgments

Support for HST programs GO-15903 and GO-15904 was provided by NASA
through an award from the Space Telescope Science Institute (STScI),
which is operated by the Association of Universities for
Research in Astronomy, Inc., under NASA contract NAS 5-26555.
All the HST data used in this paper were obtained from
the Mikulski Archive for Space Telescopes (MAST) at STScI,
with the specific observations available at
\dataset[10.17909/84nk-vt19]{http://dx.doi.org/10.17909/84nk-vt19}.

\end{document}